%% file: main.tex
\newcommand{\PaperTitle}{Non-uniformity is All You Need: Efficient and
Timely Encrypted Traffic Classification With ECHO}
\newcommand{\PaperNumber}{125}
\newcommand\ignore[1]{}

\documentclass[10pt,sigconf,letterpaper,nonacm]{acmart}

\usepackage[caption=false,font=footnotesize]{subfig}
\usepackage{verbatim}
\usepackage{tikz}
\usepackage{pgf-umlsd}
\usepackage{array}
\usepackage{rotating}
\usepackage{listings}
\usepackage{lettrine}
\usepackage{libertine}
\usepackage{courier}
\usepackage[utf8x]{inputenc}
\usepackage{float}
\usepackage{enumitem}
\usepackage{longtable}
\usepackage{mathpazo}
\usepackage{datetime}
\usepackage{booktabs}
\usepackage{multirow}
\usepackage{makecell}
\usepackage{epsfig,endnotes}
\usepackage{tabularx}
\usepackage{pifont}

\begin{document}

\title{\PaperTitle}
%\subtitle{ Under submission - do not distribute}

\author{Shilo Daum}
\affiliation{%
  \institution{The Hebrew University}
\city{Jerusalem}
\country{Israel}   
}
\email{shilo.daum@mail.huji.ac.il}

\author{Tal Shapira }
\affiliation{%
  \institution{The Hebrew University}
\city{Jerusalem}
\country{Israel}   
}
\email{talshapirala@gmail.com}

\author{Anat Bremler-Barr}
\affiliation{%
  \institution{Tel Aviv University}
\city{Tel Aviv}
\country{Israel}  }
\email{anatbr@tauex.tau.ac.il }

\author{David Hay}
\affiliation{%
  \institution{The Hebrew University}
  \city{Jerusalem}
\country{Israel}}
\additionalaffiliation{Part of this work was done while D. Hay was on a sabbatical at Princeton University, NJ, USA}
  \email{dhay@cs.huji.ac.il}

\begin{abstract}
With 95\% of Internet traffic now encrypted, an effective approach to classifying this traffic is crucial for network security and management. This paper introduces \emph{ECHO}---a novel optimization process for ML/DL-based encrypted traffic classification. \emph{ECHO} targets both classification time and memory utilization and incorporates two innovative techniques.

The first component, \emph{HO} (Hyperparameter Optimization of binnings), aims at creating efficient traffic representations. While previous research often uses representations that map packet sizes and packet arrival times to fixed-sized bins, we show that \emph{non-uniform binnings} are significantly more efficient. These non-uniform binnings are derived by employing a 
hyperparameter optimization algorithm in the training stage. \emph{HO} significantly improves accuracy given a required representation size, or, equivalently, achieves comparable accuracy using smaller representations. 

Then, we introduce \emph{EC} (Early Classification of traffic), which enables faster classification using a cascade of classifiers adapted for different exit times, where classification is based on the level of \emph{confidence}. \emph{EC} reduces the average classification latency by up to 90\%. Remarkably, this method not only maintains classification accuracy but also, in certain cases, improves it.

Using three publicly available datasets, we demonstrate that the combined method, Early Classification with Hyperparameter Optimization (\emph{ECHO}), leads to a significant improvement in classification efficiency. 
\end{abstract}

\maketitle

\input{chapters/introduction/introduction}
\input{chapters/related/related} 
\input{chapters/methods/methods}

\input{chapters/nonuniform/nonuniform2}
\input{chapters/early/early}
\input{chapters/echo/echo}
\input{chapters/results/results}
\input{chapters/conclusions/conclusions}
\label{BottomPages}
% \input{glorious}
% \input{section}
% \input{after}
% \input{another}

% Note from the CFP that this section must include a statement about
% ethical issues; papers that do not include such a statement may be
% rejected.

%%%%%%%%%%%%%%%%%%%%%%%%%%%%%%%%%%%%%%%%%%%%%%%%%%%%%%%%%%%%%%%%%%%%%%%%%%%%
% We're in the endgame now

\bibliographystyle{ACM-Reference-Format}
\bibliography{references}

\appendix
\input{chapters/methods/datasets_appendix}
\input{chapters/results/models}

%\section{Ethics}
%This paper does not raise any ethical issues.
\end{document}

%% file: chapters/introduction/introduction.tex
\section{Introduction}
\label{sec:intro}
Internet traffic classification is a critical challenge in network management, security, and optimization.
%~\cite{6644335,4534133, BUJLOW201575,4109894}. 
Traffic classification can be used for application identification~\cite{shapira2019flowpic}, traffic categorization~\cite{inproceedings, 10.1145/3447382}, identifying encryption methods like VPN or Tor~\cite{HabibiLashkari2017CharacterizationOT}, detecting malicious activities such as Distributed Denial of Service (DDoS) attacks~\cite{8171733, Sharafaldin2018TowardGA, 10.1145/3491052}, and device fingerprinting~\cite{engelberg2021classification}.

In the past, classification methods such as port-based classification (using the predefined ports associated with specific services or applications) or signature-based Deep Packet Inspection (DPI) were considered sufficient (cf.~\cite{6644335,Aceto2010PortLoadTT}). 

However, the use of random ports, shared port usage between applications, and most importantly, the widespread adoption of encryption protocols---such as TLS/SSL encryption, VPN tunneling, and anonymous communication technologies (e.g. Tor~\cite{rfc8922, 10.1007/978-3-540-31966-5_4})---have significantly hindered the ability to inspect and classify traffic based on these approaches.

Thus, new approaches were introduced to deal with encrypted traffic, mostly focused on extracting features from each flow (namely, the flow representation) and applying Machine Learning (ML) or Deep Learning (DL) methods to classify it. 
As most data is obfuscated, these methods typically use as input only the size of packets, their inter-arrival times, and their direction.

While many proposed methods claim high accuracy in classification, implementing these approaches in real-life scenarios can be challenging due to the large volume of traffic (namely, hundreds of Gb/s in large ISPs) and the huge number of concurrent connections (namely, millions of new connection per second\footnote{Recent measurements show an average of 1500 new connections per second for every 1 Gb/s of traffic, with several hundred Gb/s in a service provider network~\cite{private}.}). Thus, storing flow content, headers, or representations requires considerable memory resources. 
Moreover, the deployment of DL or large ML models demands significant computational power.

To reduce both memory and compute, flow representations often aggregate values, and each packet size (or arrival time) is mapped to a \emph{bin}. The representation then typically holds a counter for each bin, tracking the number of packets whose size (or arrival time) falls within that bin. This is most commonly done in a uniform manner with all bins being the same size. However, such coarse-grained uniform binning might fail to capture subtle nuances in traffic patterns.

\begin{figure}[tbp]
    \centering
    \subfloat[A uniform binning\label{fig:uniform}]{\includegraphics[width=0.7\linewidth]{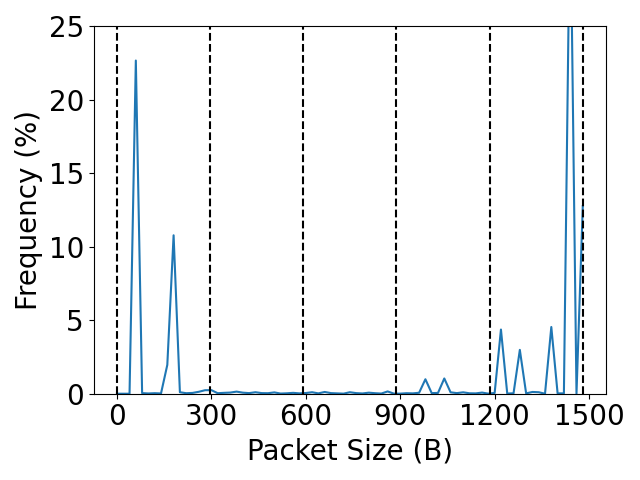}}
    \qquad
    \subfloat[A non-uniform binning\label{fig:non_uniform}]{\includegraphics[width=0.7\linewidth]{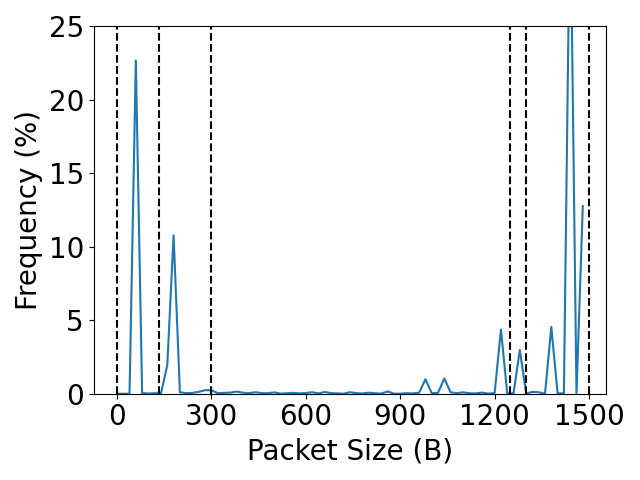}}
    \caption{A uniform and a non-uniform binning of the packet size distribution of the same example flow. Vertical dotted black lines mark the boundaries of the 5 selected bins. By using non-uniform representations, we can capture more fine-grained details of the flow.}
\label{fig:uniform_non_uniform_example}
\end{figure}

This paper first explores \emph{non-uniform binnings}, as real-world traffic data often exhibit non-uniform patterns. We introduce variable-sized bins, each representing an interval of values, allowing for a more fine-grained and adaptive approach to capturing non-uniform traffic patterns, with smaller representations. Examples of uniform and non-uniform binnings of the packet size distribution for a sampled flow are shown in Figure~\ref{fig:uniform_non_uniform_example}.

The selection of bins can be approached either manually or algorithmically. However, as there are usually multiple classes and a significant number of flows, 
manual selection of bin boundaries may be impractical and imprecise. Employing statistical methods for bin selection (or selecting only a subset of the available features) might introduce an improvement over uniform binning. Nonetheless, model-agnostic techniques might fail to capture complex relations between features. Therefore, we propose a flexible, data-aware and model-aware approach.

Innovatively, to determine the binning, we employ \emph{Hyperparameter Optimization} (\emph{HO}). In this method, we apply Bayesian optimization methods to optimize the bin boundaries in the training stage. 
Our results show that our \emph{HO} method outperforms other optimization methods, such as feature selection and statistical bin selection  across all classification tasks, by up to 15\% in accuracy. The full description of the proposed method is in Section \ref{sec:non-uniform}.

Another aspect we optimize is the collection time; that is, the time required for collecting properties from the flow until a classification is made. 
Many existing approaches classify flows only after they end, after a predefined amount of time has elapsed, or after a specific number of packets have been received. Typically, the collection time is orders of magnitude longer than the inference time, thus dominating the overall classification time. 

We notice that some flows contain indicative features useful for classification at an early stage, while others require more time for accurate classification. Thus, we introduce \emph{Early Classification} (\emph{EC}), a novel setup performing classification in multiple exit times and making a final prediction only upon reaching a certain level of \emph{confidence} in the classification. Our system is adaptive, classifying some flows at earlier stages, and waiting longer for flows with low confidence in earlier classifications. 
As our system may invoke multiple classifiers for each flow, we need to create multiple different representations, which might be prohibitive when the number of concurrent flows is large. Consequently, we present a method for creating additive traffic representations, such that the process of updating the representations between classifiers does not incur any additional memory requirements. Additionally, we propose a method that creates tradeoffs between accuracy and collection time, allowing users to tailor these parameters to their specific requirements. The \emph{EC} setup is presented in Section \ref{sec:early}.

In conclusion, this work makes two significant contributions. 
First, we introduce \emph{HO}. 
Our exploration showcases the effectiveness of this strategy, which allows us to improve the classification accuracy by 5\%--20\% in multiple classification tasks, or, alternatively, reduce memory requirements and computational overheads (as smaller classifiers are required) by 90\%.

Our second contribution is by presenting \emph{EC}, a novel method enabling early traffic classification with an adaptive classification approach. Using this method, the required time for classification can be reduced by up to an order of magnitude, with minimal decline in accuracy, and possibly even an improvement.

Finally, we combine the two methods to create \emph{ECHO} classifiers, 
achieving both an accuracy boost of up to 10\%, and a reduction in the average collection time by 95\%.

%We validate our results by performing extensive experiments and evaluations of classification accuracy and real-time deployment considerations (such as memory consumption and computational resources), using data representations and classifiers commonly used in the literature, over multiple classification tasks.

The two suggested techniques can be adapted easily to many existing approaches for traffic classification, substantially enhancing their performance.

The rest of this paper is organized as follows: We discuss related work in Section \ref{sec:related}. 
Section \ref{sec:methods} explains the networking model and defines the problem of encrypted traffic classification. 
Section \ref{sec:non-uniform} introduces \emph{HO}, where Bayesian hyperparameter optimization yields non-uniform binnings for packet sizes and arrival times. Section \ref{sec:early} outlines the Early Classification (\emph{EC}) setup. Subsequently, in Section \ref{sec:early:non_uniform}, we present the combined \emph{ECHO} approach. Finally, Section \ref{sec:res} presents the evaluation results based on three publicly-available datasets, followed by conclusions drawn in Section \ref{sec:conclusions}. 
Note that additional information, including details about our data\-sets and comparison between different models and representations, is given in the appendices.

%% file: chapters/related/related.tex
\section{Related Work}
\label{sec:related}
Encrypted traffic classification has been extensively studied in recent years (cf. \cite{survey1, survey2} and references therein). Modern classifiers predominantly utilize ML or DL techniques over representations created from the encrypted traffic. 
Commonly used representations are statistical features extracted from the traffic~\cite{article,inproceedings,HabibiLashkari2017CharacterizationOT,inproceedings_n}, time-series of the first packets of the flow\cite{10.1145/1129582.1129589,9585567}, and \emph{distributional representations}, which include \emph{dist}~\cite{engelberg2021classification}---2 pairs of distribution vectors representing packet sizes and arrival times per traffic direction---and \emph{flowpic}~\cite{10.1145/3517745.3561436, shapira2019flowpic, Finamore2023ReplicationCL}---a two dimensional representation where each element in the matrix captures the number of packet of a certain size arriving at a certain time.\footnote{While some prior research has proposed utilizing the encrypted payload of network packets for flow classification~\cite{10.1145/1080173.1080183, 8004872, ACETO2019106944, FastTraffic, Li2018ByteSN}, this approach has some inherent limitations, as properly encrypted payload data is meaningless without decryption. Therefore, it is challenging to derive meaningful insights or patterns, and the choice of encryption keys and methods can lead to shortcut learning, where the model learns to make decisions based on superficial or non-representative features, without a true understanding of the underlying data distribution (e.g., as was shown by~\cite{10.1145/3548606.3560609} with regard to~\cite{8004872}).
In light of these challenges, we focus on alternative approaches that do not rely on the content of encrypted packets.}
Notice that distributional representations often outperform other representations~\cite{Finamore2023ReplicationCL} (see further comparison results in Appendix~\ref{sec:res:models}), and therefore, are employed in this paper. 

Many methods and representations require substantial memory and compute resources, which makes them impractical for large networks.

To mitigate the memory and compute requirements, some works (e.g., ~\cite{Ede2020FlowPrintSM, 10.1145/1071690.1064220, ERTAM2017135, Shafiq2018AML, article, FAHAD20132040, 10.1145/3491052, 10.1145/3460120.3484758}), have shown how to extract certain features (e.g., flow statistics) from the traffic, or reduce feature dimensionality using feature selection. 

Non-uniform binnings (as we suggest in this work) were rarely considered in the context of flow representations. There are two notable works in this area. Barradas et al.~\cite{Barradas2021FlowLensEE} have suggested an optimization process that consists of creating uniform binning for packet size distribution and pruning bins using feature selection methods, where the final selection of features is a subset of the original \emph{uniform} bins. % and traffic in non-selected bins is ignored. 
Garcia et al.~\cite{Garcia2018EfficientDF} have proposed a statistical algorithm based on the Kolmogorov-Smirnov statistical test for selecting histogram bins, aiming at maximizing the separation between classes (measured using the Jensen-Shanon Distance). 

Our research focuses on the introduction of non-uni\-form binnings derived by a Bayesian hyperparameter optimization process.
We note that in the field of traffic classification, most previous literature regarding Bayesian hyperparameter optimization is focused on selecting hyperparameters for \emph{classifiers}~\cite{NIPS2011_86e8f7ab,pmlr-v28-bergstra13}, but not for \emph{flow representations}. \emph{HO}, our strategy for creating non-uniform representations, outperforms all previously suggested optimization methods in all explored classification tasks, and can be plugged into many of the previous works to enhance their performance and reduce their memory footprint, as detailed in Section~\ref{sec:non-uniform}. 

As for \emph{early classification},  some works have suggested using the first few packets of the flow for early classification of flows~\cite{OnTheFly, Bernaille2006EarlyAI, 10.1145/1129582.1129589,9585567,Wang2017EndtoendET, Lotfollahi2017DeepPA,8026581, Liu2019FSNetAF, Zou2018EncryptedTC}.
While for some classification tasks, the first few packets of a connection suffice, this is not always the case and can lead to less accurate classification~\cite{Finamore2023ReplicationCL}, this trend is also observable in our experiments (Appendix \ref{sec:res:models}).

%Concurrently, in the field of deep learning, some works have focused on creating deep models capable of performing early-exit classification by including internal classifiers that stop classification if a confident prediction can be made before applying all layers of the deep model~\cite{Kaya2018ShallowDeepNU, Zhou2020BERTLP}. Additionally, for general time-series data, it has been suggested to include the classification time in the training phase~\cite{8553544, Earliness}. None of these works were applied or evaluated on encrypted traffic classification tasks. 

A first step towards early classification was suggested in ~\cite{TaTic, Sivanathan2019ClassifyingID}, where two classifiers are used: the first classifier is based on simpler classification rules, while the second classifier employs more complex rules or a larger model, 
allowing for more adaptive classification sche\-mes. 
To the best of our knowledge, in the context of encrypted traffic classification, no extensive research was conducted on building \emph{confidence}-based classifiers, where a classification is made only after reaching a threshold of certainty.

%% file: chapters/methods/methods.tex
\section{Problem Statement}
\label{sec:methods}
% \subsection{Networking Model}
We first describe our networking model and define precisely the problem at hand. 

Data is transmitted over the network from a source to a destination as a \emph{flow} (namely, a timestamped sequence of \emph{packets}) over communication links. 
Typically, many flows share the same communication link. Thus, when observing a communication link, we first need to map the packets to their corresponding flows. 

We note that flows are commonly defined by a five-tuple (namely, the following five header fields in an IP packet: source and destination IP addresses, source and destination ports, and protocol)~\cite{4738466, article}. However, this definition is not rigid and can be seamlessly modified to meet different settings or requirements (e.g., in case parts of the five-tuple will be obfuscated in the future or to address a different classification task). 
%~\footnote{We note that in the literature there are alternative flow definitions (including source IP, UID, CGI, or aggregating multiple flows/connections~\cite{Roughan2004ClassofserviceMF}).} %which can be seamlessly plugged in to our system. 

This work focuses on classifying \emph{flows} that traverse the network. Thus, when a \emph{flow} (with identifier $q$) is initiated or requires classification, a \emph{flow representation} $r_q$ is generated to capture relevant properties of $q$. 
Since encryption techniques obfuscate most features, we assume \emph{only} packet sizes, arrival times, and packet directions are available. 
We aim to classify a large number (namely, millions) of flows simultaneously, implying that, at any given time, many flow representations are being collected and stored in the system.

\begin{figure}[tbp]
    \centering
    \includegraphics[width=\linewidth, trim=0cm 4cm 0cm 3cm,clip]{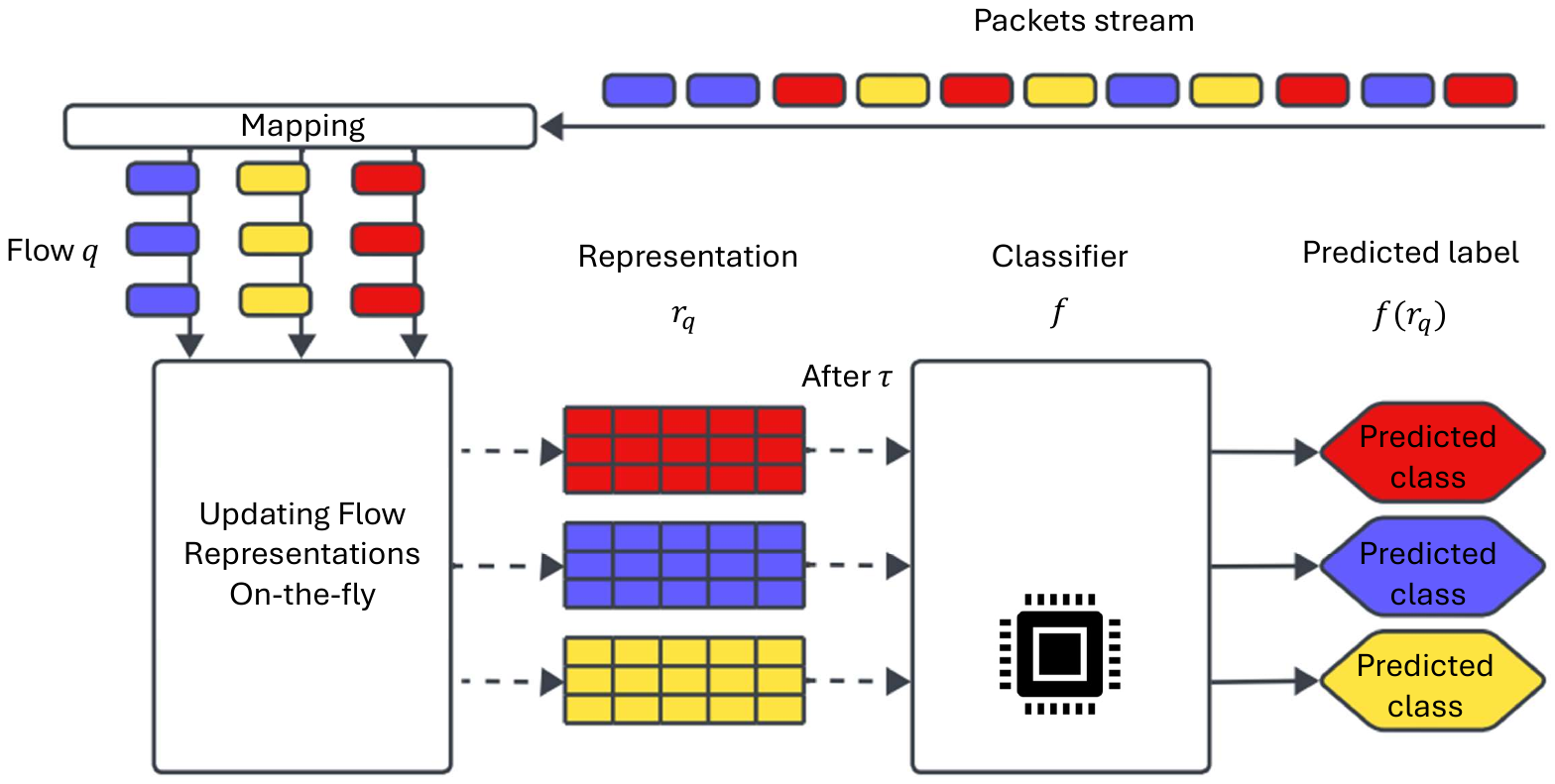}
    \caption{The key stages in the classification process. Packets are mapped by their flow (the color), and the flow representations are updated accordingly. After a predefined timeout, the flow representation is passed into a classifier to predict the class of the flow (e.g., the used application).
    Arrows represent the flow of data between the stages, whereas dotted arrows indicate memory access.
}
    \label{fig:classifier}
\end{figure}

Flow information is collected for a predefined time $\tau$, where the representation $r_q$ is built on the fly. Then, $r_q$ is passed to a classifier $f$, which predicts the label $f(r_{q})$ of flow $q$. 
Upon classification, the appropriate actions may be taken based on the predicted class, such as generating an alert, adding the information to a database, or blocking certain activities. Subsequently, the flow is marked as classified. Alternatively, time-limited classifications can be employed, where the classification remains valid for a specific duration, after which the flow should be reclassified to ensure up-to-date classification results.
Figure \ref{fig:classifier} illustrates the collection and classification process.

%% file: chapters/nonuniform/nonuniform2.tex
\section{Hyperparameter Optimization (HO) of Binnings}
\label{sec:non-uniform}
In this section, we present \emph{HO}, a novel strategy to create \emph{non-uniform binnings} that enable efficient classification. 
We have explored both non-uniformity in the packet size distribution vector and the arrival-time distribution vector. %For simplicity, we describe the creation of the binnings in the packet-size distribution vector, and the arrival-time distribution vector follows similarly (albeit with different parameters).

\emph{Binning} consists of reducing the granularity of the representation such that each original value is mapped to a \emph{bin}. Each bin represents an interval of the original values. 
Formally, a uniform representation of size $N$ for values in $[0,x)$  represents the data in $N$ bins, where bin $i\in\{1,\ldots,N\}$ is for values in $[\frac{(i-1)\cdot x}{N},\frac{i\cdot x}{N})$. 
A \emph{non-uniform} $N$-bin representation has additional degrees of freedom to choose the boundaries by defining a vector $$B=[b_0,\ldots,b_{N}],$$ such that  $0=b_0<b_{1}<\ldots<b_{N-1}<b_N=x$. In such a representation, bin $i\in\{1,\ldots,N\}$ is for values in $[b_{i-1} ,b_{i} )$.\footnote{The last bin may map also values larger than $x$ (e.g., in case of jumbo frames), and therefore, can be defined as $[b_{N-1},\infty)$.}
%Here, we aim to find the optimal non-uniform binning. 
In general, we examine two additional properties of binnings: \emph{data-aware} and \emph{model-aware} binnings. 

A binning is \emph{data-aware} if, in the creation process, it takes properties of the traffic into account. A simple example of data-aware binning is to employ \emph{Feature Selection} methods and decide not to store some of the (fixed-sized) bins, based on the data~\cite{Barradas2021FlowLensEE}.

A more sophisticated data-aware binning is to use \emph{Statistical} methods to select variable-sized bins (represented as a vector $B$) that create a separation between the distributions of the different classes of the data. Inspired by~\cite{Garcia2018EfficientDF}, we use a multiclass version of the Jensen-Shanon distance metric~\cite{jsd}, where classes are compared in an one-vs-all manner. 

A \emph{model-aware} binning, which we are first to propose, corresponds to the specific classifier that will eventually use the representations for classification. This approach allows us to tune the representation by the anticipated performance of the final \emph{model}, distinguishing it from all other binning operations (see Table~\ref{tab:freedom}). 

Orthogonal to the information available to the binning process is the way the different elements in $B$ are selected, where we distinguish between two approaches. The naive \emph{greedy} approach (as used, for example, in~\cite{Garcia2018EfficientDF}) iteratively selects a single bin boundary, maximizing the objective function given previous selections. 

\ignore{
\begin{table}[tbp] 
% \label{tab:freedom}
    \centering
   \caption{The different parameters observable by the representation optimization process}
    \small
    \begin{tabular}{|m{1.4cm}|m{1.4cm}|m{1.5cm}|m{1.5cm}|m{0.5cm}|}
    \hline
    & \textbf{Uniform} & \textbf{Feature {Selection}} & \textbf{Statistical} & \textbf{HO} \\
    \hline
    \textbf{Data-aware} & & \centering\arraybackslash  \textcolor{green!60!black}{\ding{51}} & \centering\arraybackslash  \textcolor{green!60!black}{\ding{51}} & \centering\arraybackslash \textcolor{green!60!black}{\ding{51}} \\
    \hline
    \textbf{Variable Sized Bins} & & & \centering\arraybackslash \textcolor{green!60!black}{\ding{51}} & \centering\arraybackslash \textcolor{green!60!black}{\ding{51}} \\
    \hline
    \textbf{Model-aware} & & & & \centering\arraybackslash \textcolor{green!60!black}{\ding{51}} \\
    \hline
    \end{tabular}
    \label{tab:freedom}
\end{table}
}

\begin{table}[tbp] 
    \centering
    \caption{The different parameters observable by the representation-optimization process}
    \small
    \begin{tabular}{|m{1.5cm}|m{1.5cm}|m{1.5cm}|m{1.5cm}|}
    \hline
    \textbf{} & \textbf{Data-aware} & \textbf{Variable Sized Bins} & \textbf{Model-aware} \\
    \hline
    \textbf{Uniform} & \centering\arraybackslash \textcolor{red!60!black}{\ding{55}}& \centering\arraybackslash \textcolor{red!60!black}{\ding{55}} & \centering\arraybackslash \textcolor{red!60!black}{\ding{55}}\\
    \hline
    \textbf{Feature Selection} & \centering\arraybackslash \textcolor{green!60!black}{\ding{51}} & \centering\arraybackslash  \textcolor{red!60!black}{\ding{55}}&  \centering\arraybackslash \textcolor{red!60!black}{\ding{55}} \\
    \hline
    \textbf{Statistical} & \centering\arraybackslash \textcolor{green!60!black}{\ding{51}} & \centering\arraybackslash \textcolor{green!60!black}{\ding{51}} &  \centering\arraybackslash \textcolor{red!60!black}{\ding{55}} \\
    \hline
    \textbf{HO} & \centering\arraybackslash \textcolor{green!60!black}{\ding{51}} & \centering\arraybackslash \textcolor{green!60!black}{\ding{51}} & \centering\arraybackslash \textcolor{green!60!black}{\ding{51}} \\
    \hline
    \end{tabular}
    \label{tab:freedom}
\end{table}

\emph{HO}, on the other hand, employs 
a hyperparameter optimization algorithm for binning selection.
Hyperparameter optimization, in general, refers to the process of finding the best combination of hyperparameters (configuration settings that are not learned in the training stage, but rather set before training the classifier) for a given classifier. 
In our context, the hyperparameters are the selected bins for the representation, the search space includes all possible combinations of bin boundaries, and the objective we aim to maximize in the Bayesian process is the accuracy of a trained model over the validation set.
While many hyperparameter optimization approaches exist, we focus on the Tree Parzen Estimators (TPE) method, as it shows superiority over other approaches in past experiments~\cite{NIPS2011_86e8f7ab,pmlr-v28-bergstra13}.  

%Formally, let 
%\[\mathcal{B}=\{[b_0,\ldots,b_{N}]\mid 0=b_0<b_{1}<\ldots<b_{N-1}<b_N=x\}\] 
%$\mathcal{B}$ be the set of all possible binnings.
%of size $N$ up to value $x$. 
%For each binning $B\in \mathcal{B}$, let $r^B$ be the flow representation with binning $B$, and let $r^B_q$ be the flow representation of specific flow $q$. We configure TPE to maximize the accuracy over the validation set of $f^B$; namely, \[\mbox{val\_accuracy}(f^B)=\frac{\left|q \in VS(f^B)  \mid f(r^B_q) \mbox{is correct}\right|}{\left| VS(f^B)\right|},\] where  $VS(f^B)$ is the validation set. This results in selecting the binning $\arg \max_{B\in \mathcal{B}}\{\text{val\_accuracy}(f^B)\}$.
% maybe here explanation about 

We note that, unlike~\cite{Garcia2018EfficientDF}, we apply TPE also on the \emph{Statistical} approach, where the objective function corresponds to the Jensen-Shannon distance derived from the possible binnings. While such bin selection is better than the greedy approach, it is still inferior to \emph{HO} as the \emph{Statistical} approach is not model-aware.

For the experiments in this work, we use $k$-fold cross-validation (\emph{KFCV})~\cite{Kohavi1995ASO}.
Therefore, to validate the TPE process, we use nested cross-validation (NCV)~\cite{10.5555/1756006.1859921}, an extension of \emph{KFCV} for hyperparameter optimization. 
NCV consists of an outer loop for standard \emph{KFCV}, selecting training and test sets in each iteration. In the inner loop, we use the training data to iteratively select optimal hyperparameters through another \emph{KFCV} process. We use $k=5$ in both the outer and inner loops, implying that for each evaluation in the inner loop, $64\%$ of the flows are used for training, $16\%$ for validating the hyperparameters by calculating the objective function, and $20\%$ for testing (in the outer loop). It is important to note that different iterations of the outer loop might yield different hyperparameter values. 

Notice that in our settings, the optimization process is done offline on labeled data, and therefore, has fewer constraints regarding computational power or latency. 

Once the optimization process finishes and an optimal result is achieved, the real-time classifier is adapted and uses the optimal hyperparameters (namely, the bin boundaries) achieved by the offline optimizer. 

After selecting the bin boundaries, to create our representations without any computational overhead, we create a mapping, using a small direct-access array of size $x$,  between every possible packet size (or discretized arrival time) to its bin index; this implies that the process of building our representations (which is done online, packet by packet) does not incur additional computations (only one memory access). 

To capture changes in traffic patterns or in the underlying applications, we propose to run \emph{HO} periodically, upon degradation of accuracy, or manually if a change in the network condition is known. The process should run in the background without interfering with ongoing classifications.

The results achieved using \emph{HO}, as well as a comprehensive comparison to the alternative methods, are shown in Section \ref{sec:res:non_uniform}. 

%% file: chapters/early/early.tex
\section{Early Classification (\emph{EC}) of Traffic}
\label{sec:early}
In this section, we will describe the system setup enabling \emph{early traffic classification}. \emph{EC} not only improves the Quality of Service (QoS) and the network security but also saves memory requirements, as flow representations are stored for shorter periods. 

This method is based on the assumption that some flows have indicative features at earlier stages and there is no need to wait for a long time to reach an accurate classification. Therefore, in the proposed setup, classification occurs upon reaching a sufficient \emph{confidence} level, without the need to wait several seconds and collect the full traffic representation. For assessing models' \emph{confidence}, we exploit the fact that many classifiers output the predicted results as a vector $f'(r) = u$, where $u$ consists of the predicted probability that a sample is in each class. The predicted class for some sample is then  $f(r) = \arg \max_{i =1}^{C} f'(r)$. For EC, we define the \emph{confidence} of the classification, denoted by $\beta_q$, as the maximal argument in the resulting vector; namely,
$$\beta_q = \max_{i=1}^{C} f'(r_q).$$
We note that this approach can be replaced with several different methods for assessing model confidence, as discussed by \cite{confidence}.
 
The core idea behind \emph{EC} is that classification occurs in multiple exit times $\tau_1,\ldots,\tau_{max}$, and the final prediction is made when the \emph{confidence} value exceeds a predefined threshold $\beta$, or if the maximal time $\tau_{max}$ has elapsed. 

For employing an efficient \emph{EC} system, we first train (offline) different classifiers $f_1,\ldots,f_{max}$ for multiple durations in a logarithmic scale (namely, $\tau_{i+1}=2\cdot \tau_{i}$) up to the maximal duration $\tau_{max}$ selected for the classification task.

\begin{figure}[tb]
    \centering
    \includegraphics[trim=120pt 30pt 150pt 30pt, clip, width=0.95\linewidth %height=0.37\linewidth
    ]{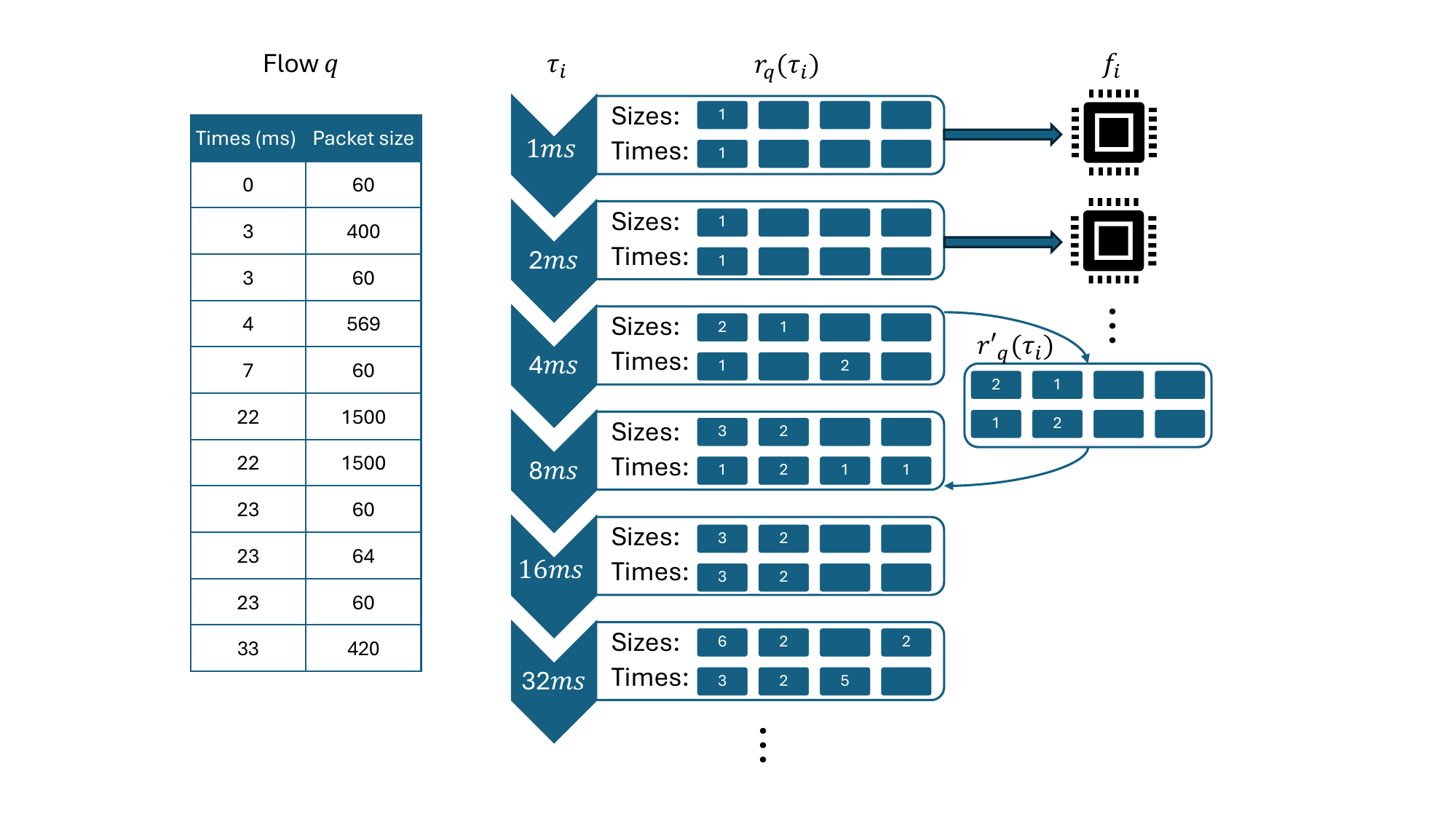}
    \caption{An illustration of the iterative process of creating the representations for the different time scopes. On the left is an example flow with the arrival times and packet sizes of the first 11 packets. The packet-size boundaries vector is $[0,375,750,1125,1500]$, the arrival time boundaries vector is dependant on the time scope: $[0,\frac{\tau_i}{4},\frac{\tau_i}{2},\frac{3\tau_i}{4},\tau_i]$.
    $r^{'}_{q}(\tau_i)$ is the updated representation before adding the packets in the next time interval. %Notice that this illustration only shows one side of the traffic but in practice, the representation includes counter vectors for the two sides of the traffic. 
    }
    \label{fig:early_process}
\end{figure}

In the online phase, we iteratively build the traffic representations. First, we build the distribution vectors with regard to the first exit time and feed it into the first classifier. If the sample is not classified with sufficient \emph{confidence}, we keep the packet-sizes distribution vector unchanged and keep collecting packets to it. 
For the arrival-times distribution vector, in order to adjust the representation to the next classifier, 
we first merge each bin $2i-1$ with its adjacent bin $2i$, leaving the second half of each vector empty. Then, 
we count arriving packets into the second halves of these vectors, where each bin corresponds to the time that passed since the beginning of the collection process.
Formally, for $d({\tau_i})$ being a distribution vector of the arrival times for the exit time $\tau_i$ (for a single side of the traffic), the values in the updated vector $d({\tau_{i+1}})$ for the next exit time will be: 
%$$r^{'}_{k}(\tau_i)[sizes0,1] = r_{k}(\tau_i)[sizes0,1],$$ and
% $$r^{'}_{k}(\tau_i)[times0,1]_j = \begin{cases}
%     r_{k}(\tau_i)[times0,1]_{2j-1}+r_{k}(\tau_i)[times0,1]_{2j} &: j\leq \frac{N}{2}\\
%     0 &: \text{else}\\
% \end{cases}.$$
$$d({\tau_{i+1}})_j = \begin{cases}
    d({\tau_{i}})_{2j-1}+d({\tau_{i}})_{2j} &: j\leq \frac{N}{2}\\
    \#(\text{packets in }[\frac{j-1}{N}\cdot 2\tau_{i}, \frac{j}{N}\cdot 2\tau_{i})) &: \text{else}\\
\end{cases}.$$
Using this method, no additional memory is required, and the representations are built and adjusted to the different classifiers on the fly. This process is illustrated in Figure \ref{fig:early_process}. 

To minimize the negative effect this process might have on classification, we select the \emph{confidence} threshold $\beta$ as the accuracy of a baseline model that classifies all flow after $\tau_{max}$.

Surprisingly, this scheme not only reduces the average classification time but in some cases also improves the overall classification accuracy. This might occur due to the multiple exit options, allowing the models to identify timely behaviors that are more discernible in earlier classifiers. 

Additionally, we explore the tradeoffs between the average collection time and the overall classification accuracy by altering the classification threshold $\beta$
with small $\alpha$ values, making early predictions if the confidence is higher than $\beta - \alpha$.

% As shown in Section \ref{sec:res:throughput}, the classification throughput of LR classifiers on the tested machine is at about millions of flows per second. Therefore, the classification throughput (i.e., the computational power) is not a bottleneck in this case. 
As shown in Appendix \ref{sec:res:throughput}, the classification throughput highly depends on the selected classifier and the representation size. Therefore, if \emph{EC} is deployed on edge devices, at high network rates, or if a strong classifier is selected, there will be a tradeoff between the ability to perform early classification and the rate of classification, as this method requires (possibly) employing multiple classifiers on each flow until it reaches the final prediction. Additionally, as the collection time is significantly higher than the inference time of our classifiers, we neglect the inference time in these comparisons. \footnote{See Appendix \ref{sec:res:throughput}}.

The results exploring \emph{EC} setups are shown in Section \ref{sec:res:early}.

%% file: chapters/echo/echo.tex
\section{Early Classification with Hyperparameter Optimization (ECHO)}
\label{sec:early:non_uniform}
As we observed that \emph{HO} improves the performance of models, we wish to create those representations for \emph{EC} setups as well. 
For the packet-size distribution vector, we adopt the non-uniform bins of an optimized \emph{HO} model (a single binning for all classifiers). 
However, as the arrival time distribution vector are updated in a particular way between EC classifiers (recall Figure~\ref{fig:early_process}),   customized arrival-times bins cannot be used. The \emph{EC} setup employing \emph{HO} on the packet size dimension with uniform time bins is denoted by \emph{ECHO}. 

Yet, we do propose a non-uniform alternative for the arrival time bins, while still maintaining the ability to update the representations in real-time. In this alternative setup, denoted by \emph{EC Log}, we will use \emph{logarithmic bins}, that is, for every exit time $\tau_i$ and $N$ bins with boundaries $B = [b_0,\ldots, b_N]$, where $b_0=0s,\: b_N=\tau_i$, and the other boundaries ($0<j<N$):
$$b_j = {\tau_i}\cdot {2^{-j}}.$$
Using this selection, if the different exit times are defined logarithmically ($\tau_{i+1} =2\cdot \tau_{i}$), to update the arrival-times vector distribution we simply sum the two values with the smallest indexes, and shift all values one index below.
Formally, for $d({\tau_{i}})$ being a single distribution vector of arrival times created at time ${\tau_{i}}$, $d({\tau_{i+1}})$ will be the updated vector for the next classifier:
$$d({\tau_{i+1}})_j = \begin{cases}
    d({\tau_{i}})_{1}+d({\tau_{i}})_{2} &: j=1\\
    \#(\text{packets in } [\tau_{i}, \tau_{i+1})) &: j=N\\
    d({\tau_{i}})_{j-1} &: \text{else}\\
\end{cases}.$$
% $$r^{'}_{k}(\tau_i)[times0,1]_j = \begin{cases}
%     r_{k}(\tau_i)[times0,1]_{1}+r_{k}(\tau_i)[times0,1]_{2} & :j=1\\
%     0 &: j=N\\
%     r_{k}(\tau_i)[times0,1]_{j-1} &:\text{else}\\
% \end{cases}.$$
In Section \ref{sec:res:early_non_uniform}, we have evaluated different selections of packet-size bins for the \emph{ECHO} setup, and explored the proposed method for the logarithmic arrival-times bins (\emph{EC Log}). 

%% file: chapters/results/results.tex
\section{Experimental Results}
\label{sec:res}
In this section, we provide experimental results investigating the \emph{HO} strategy for creating efficient traffic representations (Section \ref{sec:non-uniform}), the \emph{EC} method for timely traffic classification (Section \ref{sec:early}) and the \emph{ECHO} approach that combines the two strategies (Section \ref{sec:early:non_uniform}). We also demonstrate their superiority over alternatives that were previously proposed in the literature.

Our results are structured as follows:
First, we examine the effectiveness of \emph{HO} in traffic classification (Section \ref{sec:res:non_uniform}). Here, we analyze how the representations created using \emph{HO} impact classification accuracy and memory utilization. Our investigation highlights the advantages of \emph{HO} compared to traditional uniform binning and alternative approaches, and provides in-depth insights into the reasons behind the success of the proposed method. % We aim to unravel the explanatory factors that contribute to their superior performance.

Section \ref{sec:res:early} explores \emph{EC} methods and how they affect the accuracy and reduce the average collection time of the system. 

Finally, we combine the \emph{HO} method with \emph{EC} to create efficient classifiers, with high accuracy and with minimal time required for classification (Section \ref{sec:res:early_non_uniform}).

In this work, we use the Logistic Regression ($LR$)~\cite{doi:10.1073/pnas.6.6.275} classifier for building the models. $LR$ classifiers offer the highest classification throughput compared to other explored classifiers (a detailed comparison in Appendix \ref{sec:res:models}), therefore enabling real-time classification for large volumes of traffic. However, we note that in our experiments, similar trends were observed for different ML and DL classifiers.

We use three publicly-available datasets~\cite{rezaei2020achieve,inproceedings, HabibiLashkari2017CharacterizationOT,NAAS2023108945}, where classification tasks include application identification (e.g., distinguishing between YouTube and Google Music), encryption identification (e.g., distinguishing between different types of VPN), and  traffic categorization (e.g. distinguishing between video and VoIP). The different classification tasks we tackle are presented in Table \ref{tab:tasks}, and a thorough description of the datasets, including our preprocessing stage, is in Appendix \ref{sec:datasets}.

It is important to note that our preprocessing ensures that the datasets are balanced across all classification tasks, and the difference between accuracy (percentage of correctly classified samples over the test set) and alternative utility metrics (e.g. f1 score, precision, recall) is negligible. Consequently, we evaluate our models and representations based solely on the achieved accuracy using \emph{$k$-fold} Cross Validation (KFCV)~\cite{Kohavi1995ASO} with $k=5$.

\begin{table}[tb]
\centering
\caption{Different classification tasks we explore. Classification task types are denoted ``Apps''  for application identification, 
``Encryption'' for encryption identification, and ``Categories'' for traffic categorization. Additional details appear in Appendix \ref{sec:datasets}.}
\label{tab:tasks}
\footnotesize
{
\begin{tabular}{|l|l|c|l|}
\hline
 \textbf{Dataset Name} & \textbf{Task Name} & \textbf{\# Classes} & \textbf{Type} \\ \hline
Quic~\cite{rezaei2020achieve} & Quic Applications &  5 & Apps \\ \hline
\multirow{ 3}{*}{\shortstack[l]{ISCX combined \\ \cite{inproceedings, HabibiLashkari2017CharacterizationOT}}} & ISCX Applications  & 4 & Apps \\ \cline{2-4}
 &ISCX Encryption  & 3 & Encryption \\ \cline{2-4}
 & ISCX Categories   & 4 & Categories \\ \hline
\multirow{ 2}{*}{\shortstack[l]{VPN Services\\  \cite{NAAS2023108945}}} & VPN Services  & 6 & Encryption \\ \cline{2-4}
&Streaming or Not &  2 & Categories \\ \hline
\end{tabular}
}
\end{table}

\subsection{Hyperparameter Optimization
(HO) of Binnings}
\label{sec:res:non_uniform}

As observed in our experiments, larger representations improve classification accuracy, but come at the cost of a higher memory footprint and usually more computational requirements. 

We compare several methods for selecting binnings over the packet size dimension, including our suggested novel strategy:
\begin{description}[leftmargin=0cm,labelindent=0cm,parsep=0.5em]
\item[\emph{Uniform}.]  All bins are of equal size; namely,  $b_i=\frac{i \cdot x}{S}$, where $S$ is the number of bins.

 \item[\emph{Feature Selection}.] 
Inspired by~\cite{Barradas2021FlowLensEE}, we employ feature-selection methods on larger uniform representations, to create a smaller pruned representation. Given the required number of bins $N$, we apply feature selection on multiple uniform representations with different original sizes: $N' \in \{10,20,50,100,200,500,1500\},$ and select the model with the highest achieved accuracy. This representation still uses fixed-size bins, however, it is \emph{data-aware} and selects features with respect to their statistical significance. 

\item[\emph{Statistical}.] 
As explained in Section~\ref{sec:non-uniform}, and inspired by~\cite{Garcia2018EfficientDF}, we use the TPE method for optimization of binning, where the objective function is to maximize a multiclass version of the Jensen-Shanon distance metric~\cite{jsd}. 
%providing the maximal  statistical seperation between the distributions of the different classes. Recall that
%to create the maximal separation between the packet size distributions of different classes. Specifically, we use the Jensen-Shanon distance metric~\cite{jsd} (extended to multiclass using one-vs-all). 
%
This method maximizes the statistical separation between the different classes independently of the selected classifier.

\item[\emph{HO Sizes}.] 
Our novel strategy employed on the packet size distribution, using the TPE method with the objective being the accuracy of a trained model, as explained in detail in Section \ref{sec:non-uniform}.
\end{description}
\begin{figure*}[btp]
    \centering
    \subfloat[Quic Applications\label{fig:quic_apps_compare_opt}]{\includegraphics[width=0.3\linewidth]{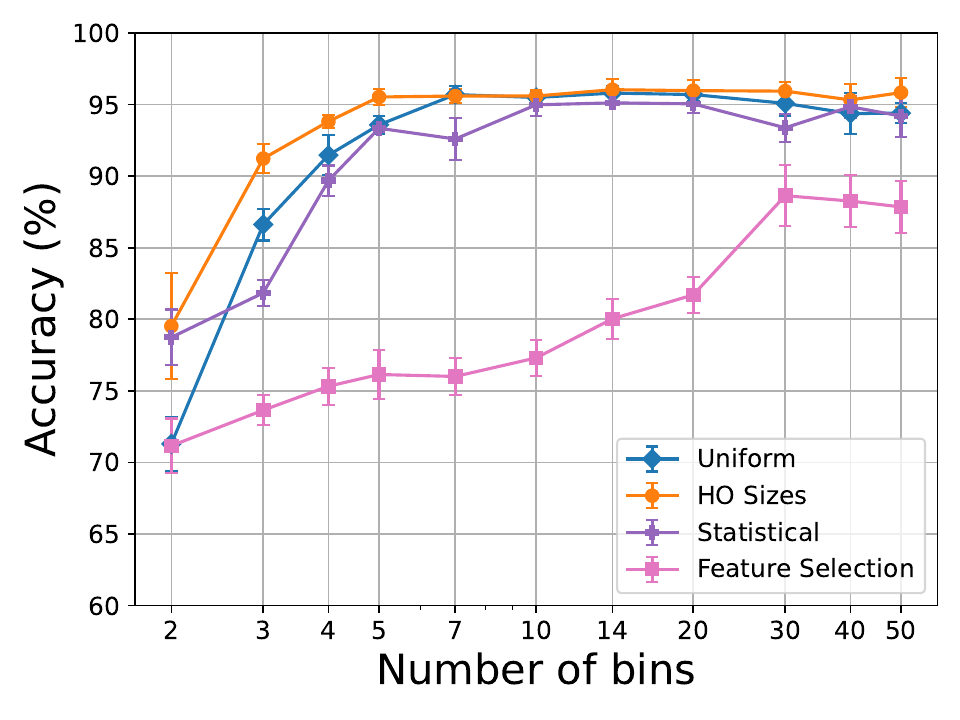}}
    \qquad
    \subfloat[ISCX Applications\label{fig:iscx_apps_compare_opt}]{\includegraphics[width=0.3\linewidth]{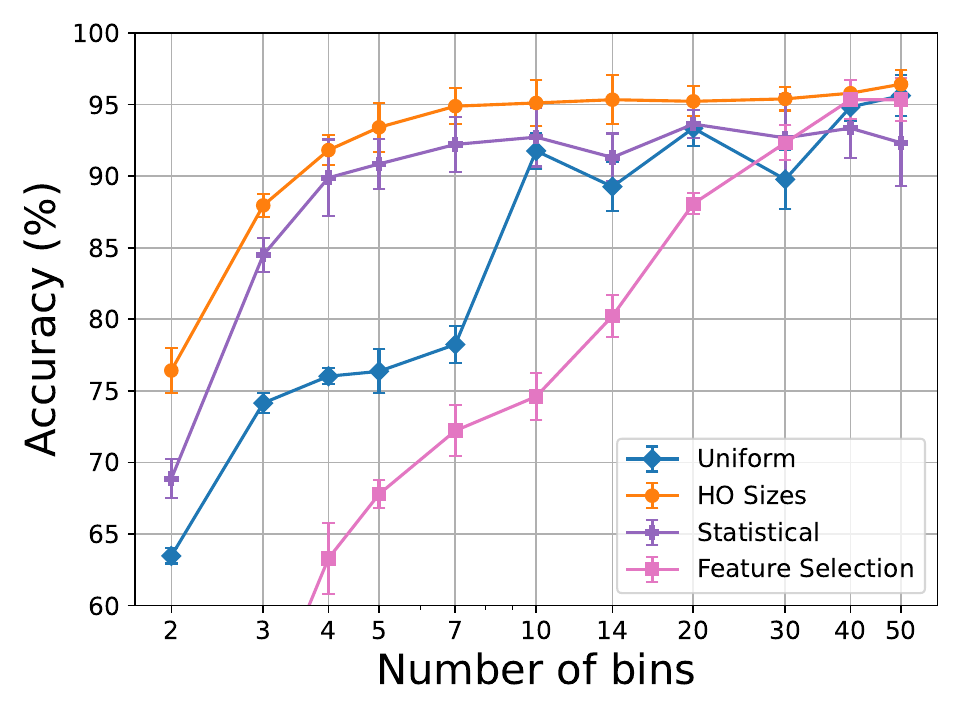}}
    \qquad
    \subfloat[ISCX Encryption\label{fig:iscx_encryption_compare_opt}]{\includegraphics[width=0.3\linewidth]{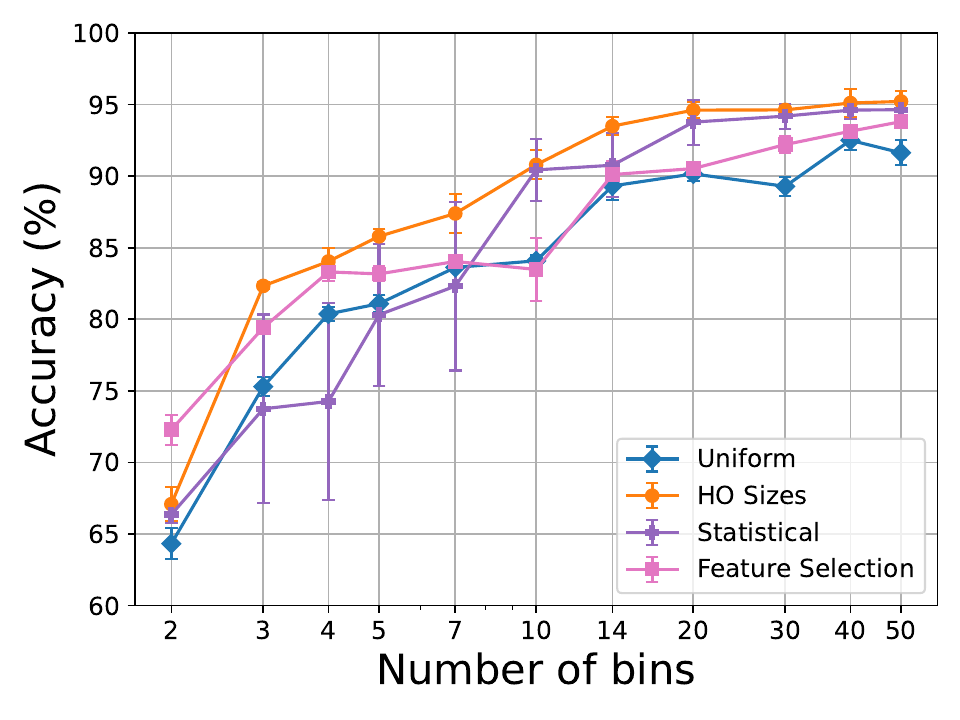}}
    \qquad
    \subfloat[ISCX Categories\label{fig:iscx_categories_compare_opt}]{\includegraphics[width=0.3\linewidth]{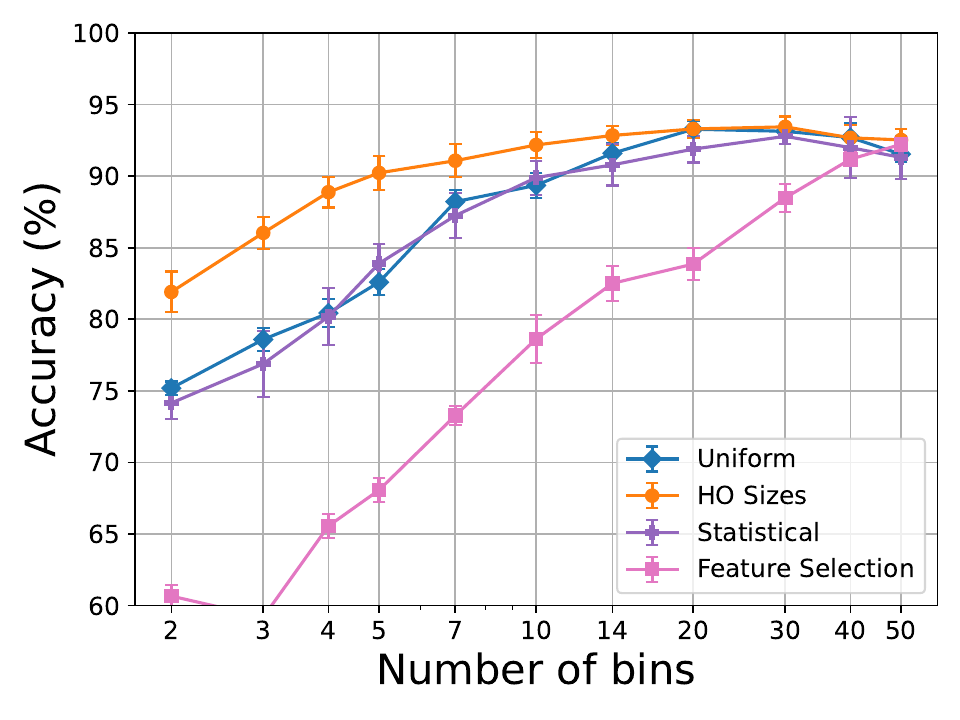}}
    \qquad
    \subfloat[VPN Services\label{fig:vpn_service_compare_opt}]{\includegraphics[width=0.3\linewidth]{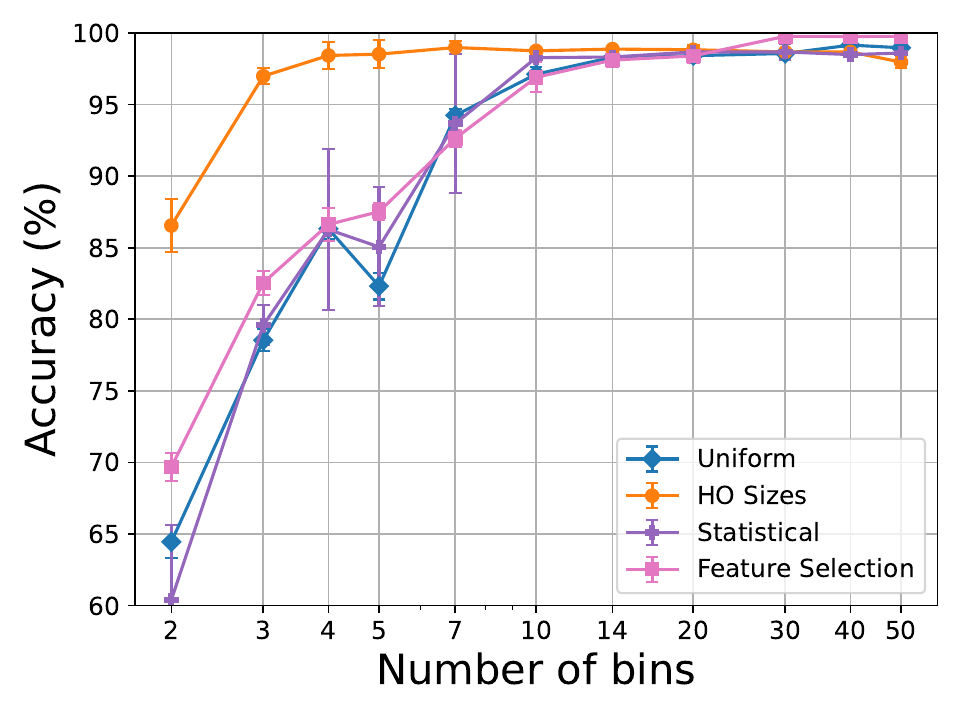}}
    \qquad
    \subfloat[Streaming or Not\label{fig:strm_or_not_compare_opt}]{\includegraphics[width=0.3\linewidth]{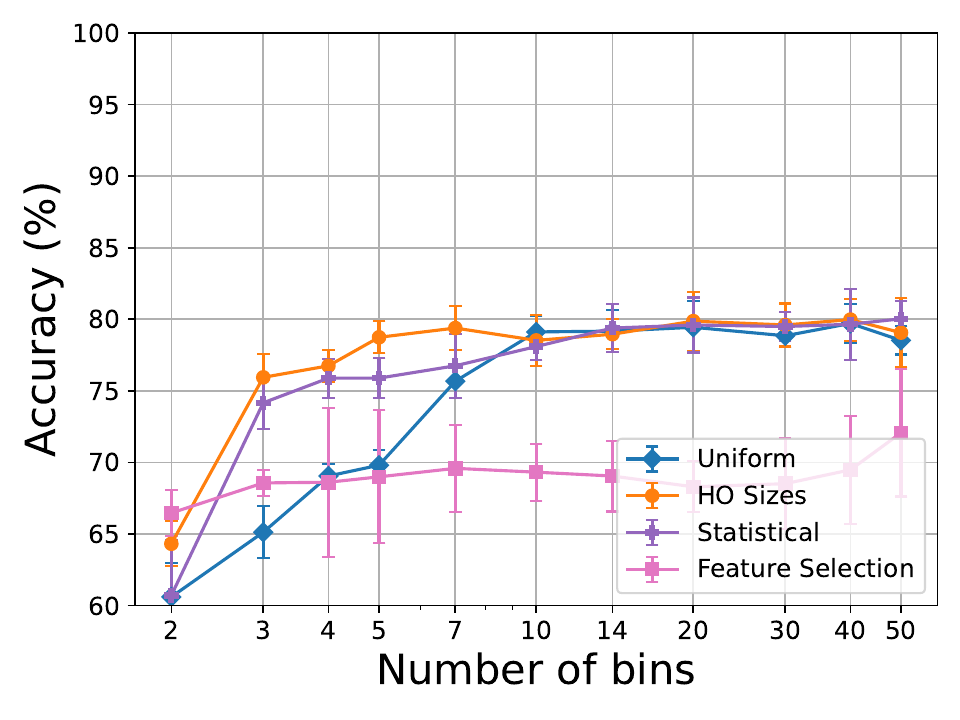}}

    \caption{Accuracy as a function of the number of bins, for different binning strategies on the packet size dimension. The time dimension binning is uniform in all experiments.}
    \label{fig:compare_opt}
\end{figure*}
%For our approach, as well as for the statistical approach, we employ a Bayesian TPE sampling algorithm to enable efficient sampling from a multidimensional space of parameters (as presented in Section \ref{sec:non-uniform}), the difference between the two methods lies in the maximization function, wherein the statistical method, this function is based solely on the distribution of packet sizes, with no regard to a specific model or classification task, while in our method, we use the trained model accuracy as the maximization objective. 
Throughout all our experiments, we set the number of iterations of the TPE method to 200, as we observed convergence in this iterative process.
Note that we have also explored greedy sampling methods for hyperparameter selection, where we progressively built the hyperparameter configuration by selecting the option in each subsequent bin that maximized the objective function given the previously chosen options. 
However, as greedy selection methods do not have as many degrees of freedom, and showed no noticeable advantage over the TPE method, we focus our discussion here on the TPE.

%We suggest a novel method for creating representations that capture the fine-grained details of the flows without a high memory footprint.
The results are shown in Figure \ref{fig:compare_opt}.
Evidently, the \emph{HO Sizes} approach is able to capture the nuances and patterns in smaller representations, leading to improved classification accuracy throughout all classification tasks. 
Additionally, statistical optimization (\emph{Statistical}) shows an advantage over uniform representations; however, this trend is not consistent for all tasks and sizes.
Finally, the \emph{Feature Selection} method introduces a degradation in accuracy compared to uniform representations. This might be caused by the loss of information in the feature selection process, in the inherent limitation of selecting only sampled features for a single uniform representation (of larger size). 

\begin{figure*}[btp]
    \centering
    \subfloat[Quic Applications\label{fig:quic_apps_sizes_times_bins}]{\includegraphics[width=0.3\linewidth]{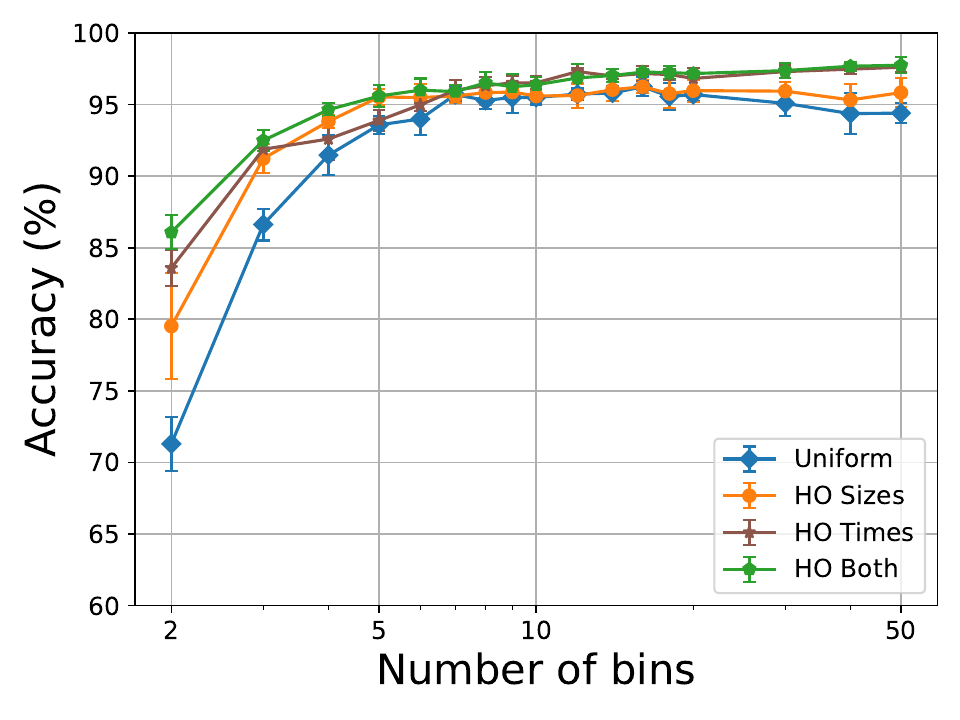}}
    \qquad
    \subfloat[ISCX Applications\label{fig:iscx_apps_sizes_times_bins}]{\includegraphics[width=0.3\linewidth]{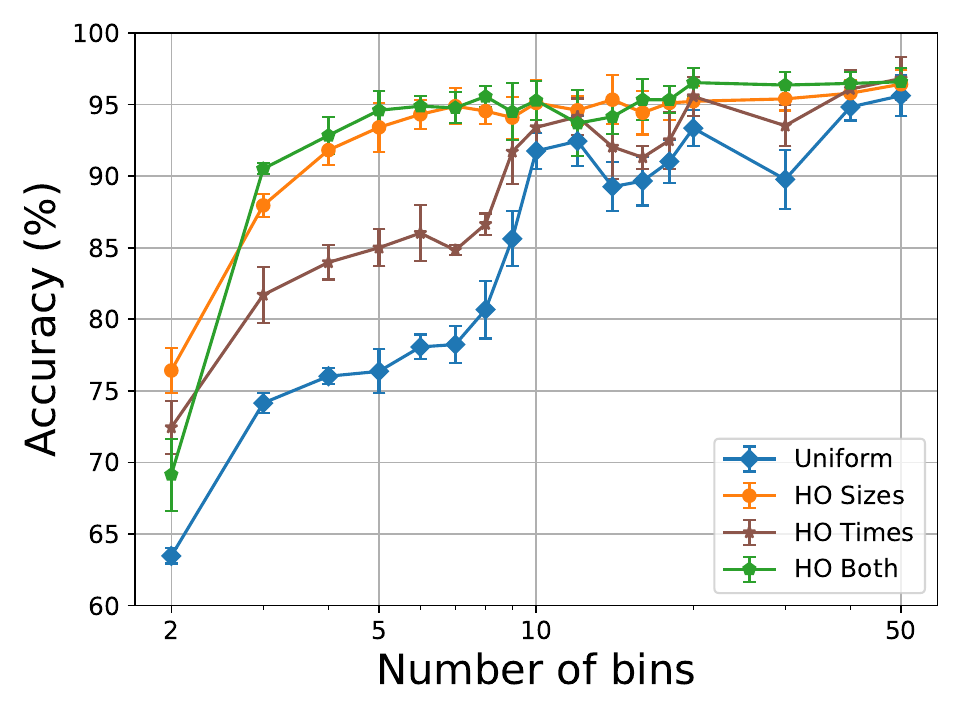}}
    \qquad
    \subfloat[ISCX Encryption\label{fig:iscx_encryption_sizes_times_bins}]{\includegraphics[width=0.3\linewidth]{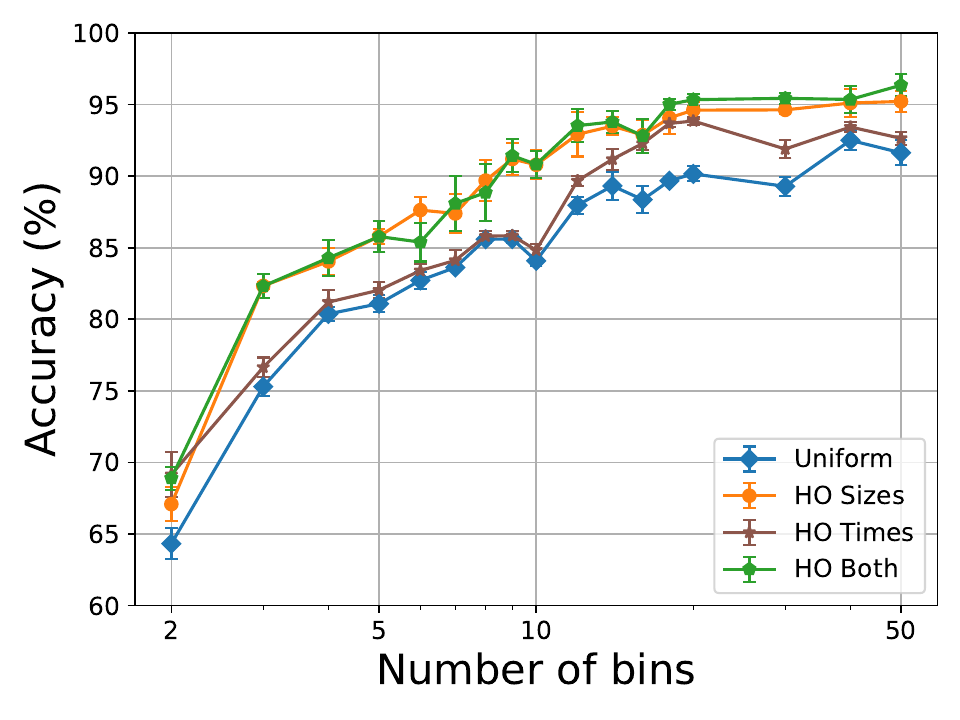}}
    \qquad
    \subfloat[ISCX Categories\label{fig:iscx_categories_sizes_times_bins}]{\includegraphics[width=0.3\linewidth]{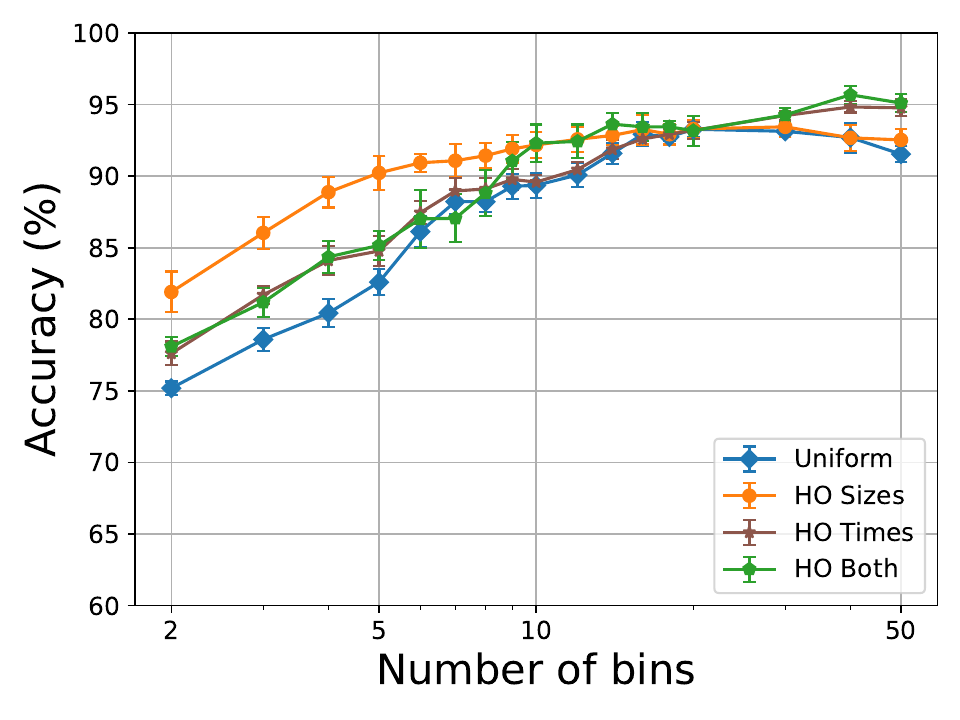}}
    \qquad
    \subfloat[VPN Services\label{fig:vpn_service_sizes_times_bins}]{\includegraphics[width=0.3\linewidth]{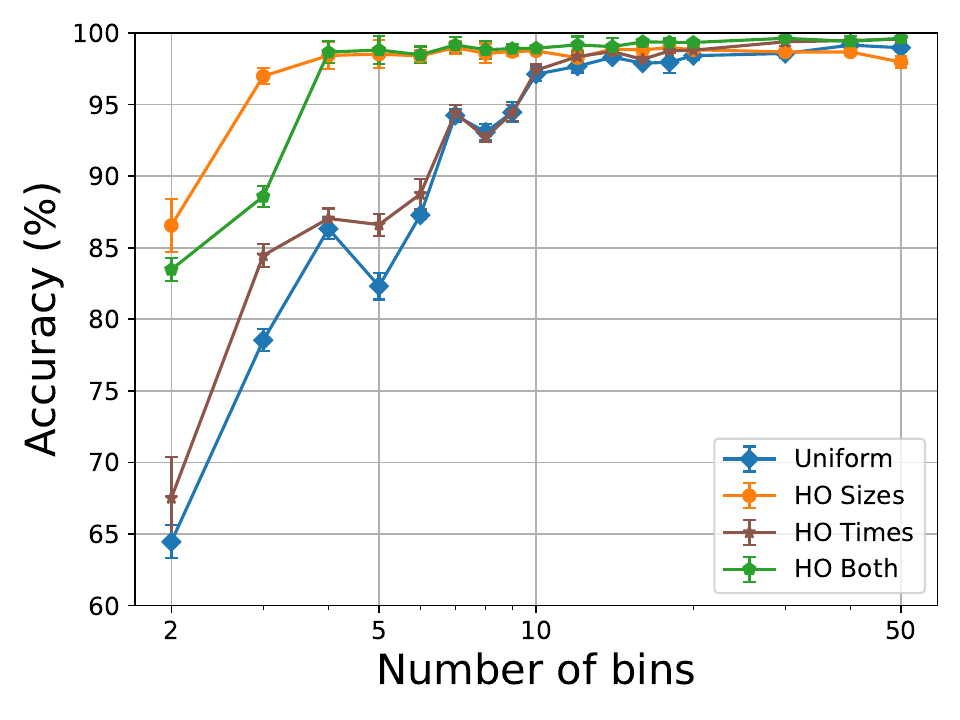}}
    \qquad
    \subfloat[Streaming or Not\label{fig:strm_or_not_sizes_times_bins}]{\includegraphics[width=0.3\linewidth]{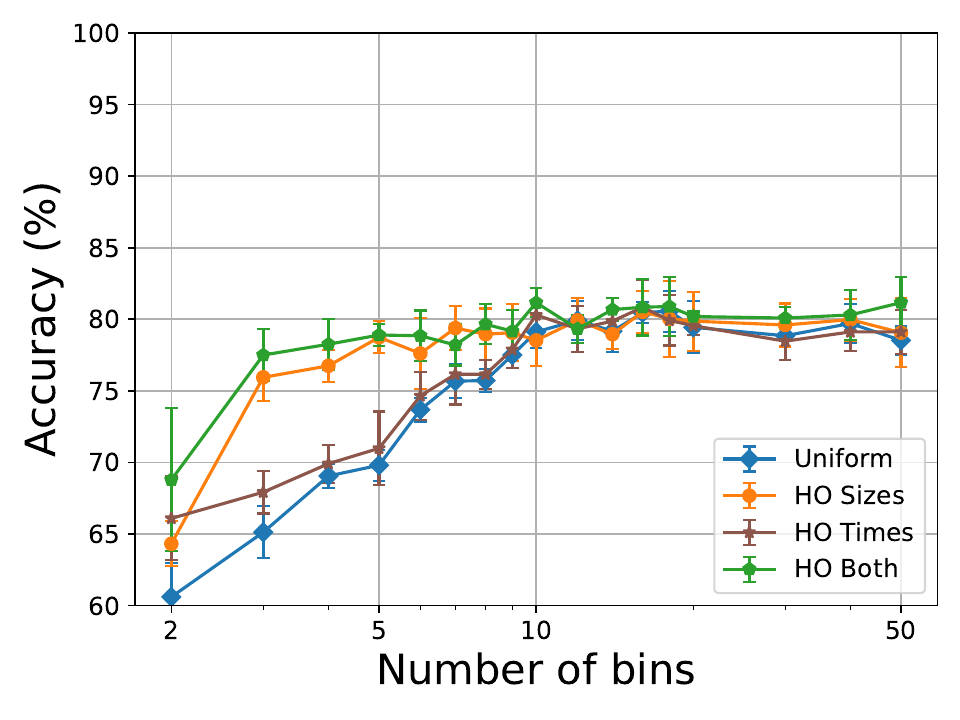}}

    \caption{Accuracy as a function of the number of bins, comparing \emph{HO} on the different dimensions, as well as on both dimensions.}
    \label{fig:sizes_times_bins}
\end{figure*}

We note that introducing non-uniformity in the packet sizes dimension leads to a significant enhancement in the model's accuracy, particularly evident when employing the \emph{HO Sizes} approach. We further explore non-uniformity by introducing non-uniformity in the arrival-times dimension.
We compare the following representations: \emph{Uniform} representations, \emph{HO Sizes} - that use non-uniform bins for the sizes dimension and uniform bins for the arrival-times dimension (as in Figure \ref{fig:compare_opt}), \emph{HO Times} - that use non-uniform bins for the arrival time dimension and uniform bins for the sizes dimension, and \emph{HO Both} that include optimized binning for both dimensions (with different parameters for each dimension).
Results are shown in Figure \ref{fig:sizes_times_bins}. We see that by itself, non-uniformity in the arrival-times dimension shows some advantage over uniform representations. However, this improvement is limited, and when combined with non-uniformity in the packet-sizes dimension, there is little to no advantage in model accuracy compared to \emph{HO Sizes}.

\subsubsection{Explainability}
\label{sec:explain}
\begin{figure*}[htb]
\includegraphics[width=0.6\linewidth]
{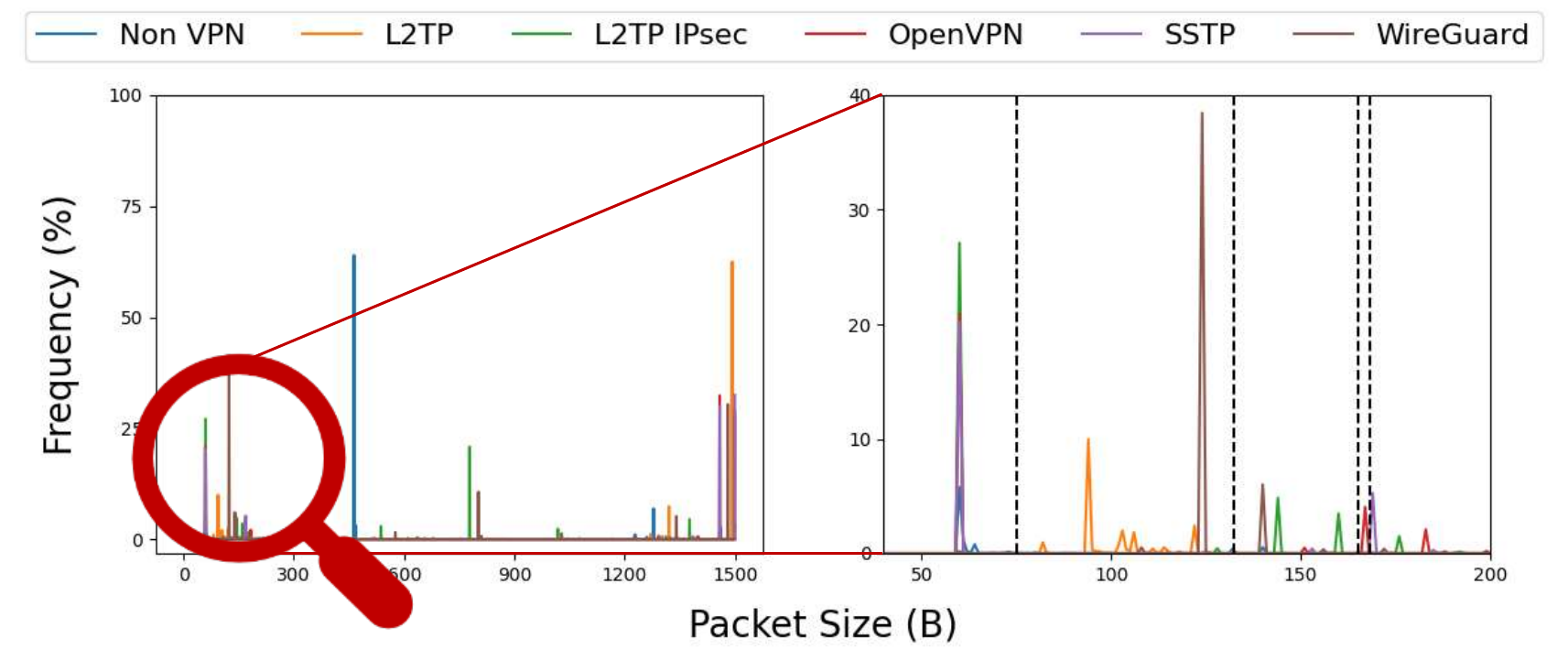}
        \caption{(a) The entire packet sizes distribution of the VPN dataset with respect to VPN labels; (b) The histogram zoomed in a specific range with the bin boundaries selected by the \emph{HO Sizes} strategy (for the VPN Services classification task with 5 bins) marked in black. We can see how the selection of bins correlates to the discriminative features of some classes.}
        \label{fig:explain_bin}
\end{figure*}

\begin{figure}[tbp]
    \centering    \includegraphics[width=0.7\linewidth]{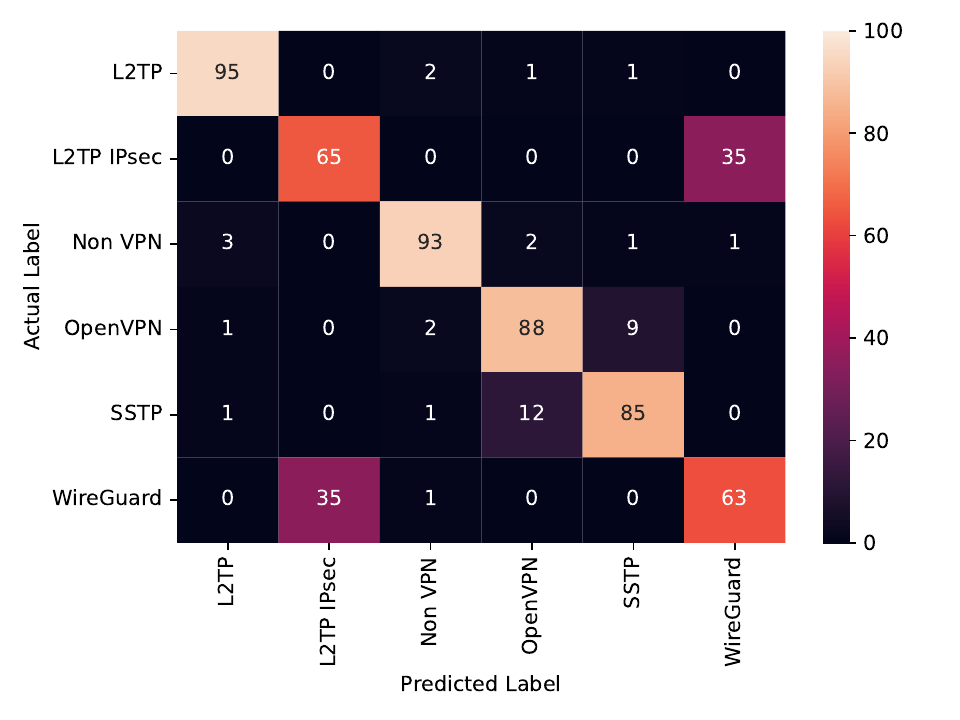}
    \caption{A confusion matrix of an $LR$ classifier using flow representations with 5 uniform bins. Each entry denotes the proportion of flows belonging to a certain class that have been predicted to fall into another class. Notable misclassifications are between L2TP IPsec and WireGuard, as well as between OpenVPN and SSTP.}

    \label{fig:dist(5)_conf}
\end{figure}

An important property of our binning strategy is its \emph{explainability}. This is done by looking at the distribution of packet sizes (or arrival times) for each class, and observing what information is given by the new bin boundaries.
%We want to emphasize the \emph{explainability} of the suggested methods for optimization. After the process of tuning the bin boundaries, we can plot the distribution of packets for each class, and see what information is given by the new bin boundaries. 
For example, for the VPN Services task over 5 bins, Figure \ref{fig:explain_bin} depicts the selected bin boundaries of one of the runs of \emph{HO Sizes}, along with the packet sizes distribution of each class. It is easy to see that bin boundaries are correlated to the data distribution. Specifically, notice that one of the boundaries is at 168 bytes. 
% (in fact, many of our TPE-generated non-uniform representations for this task included this boundary). 
In retrospect, this appears to distinguish between typical SSTP packets and typical OpenVPN packets. Similarly, the bin boundary of 76 bytes appears to distinguish between L2TP IPsec traffic (which has many smaller packets) and WireGuard traffic (with bigger packets).
These boundaries demonstrate the strength of the \emph{HO} approach, as coarse-grained uniform representations commonly group these values together resulting in misclassifications; for example, Figure \ref{fig:dist(5)_conf} shows a confusion matrix of a classifier using coarse a flow representation with 5 uniform bins. 
We can see that the classes that are misclassified are the classes for which the boundaries selected by \emph{HO Sizes} provide the necessary separation. An optimized representation employing an $LR$ classifier with the same number of bins achieved an accuracy of $98\%$, in contrast to all other binning methods achieving accuracy lower than $90\%$, as the \emph{Uniform} and \emph{Feature Selection} approaches are limited to bins of equal size, and the bins selected by the \emph{Statistical} approach were correlated to the overall data distribution but were unable to capture small nuances that affected the final classification performance. 
\subsection{Early Classification (EC) of Traffic}
\label{sec:res:early}
This section explores our EC setup, which is based on training multiple classifiers $f_1,\ldots,f_{max}$ for multiple exit times $\tau_1,\ldots,\tau_{max}$. When a classifier $f_i$ ($1\leq i<max$) achieves a \emph{confidence} value 
larger than a threshold $\beta$, the classification stops and the label predicted by $f_i$ is reported. Otherwise, after the maximal exit time $\tau_{max}$, the label predicted by $f_{max}$ is reported (regardless of its confidence). A detailed explanation can be found in Section \ref{sec:early}. In this section, for ease of presentation, we used a pseudo-logarithmic scale $(1ms,2ms,5ms,10ms \ldots)$.

First, to gain a sense of the behavior of the confidence for the different classifiers, we train the classifiers $f_1,\ldots,f_{max}$ and checked their classification accuracy and coverage (the proportion of instances for which the classifier provides a prediction) for possible \emph{confidence} thresholds.

\begin{figure}[tbp]
    \centering\includegraphics[width=\linewidth]{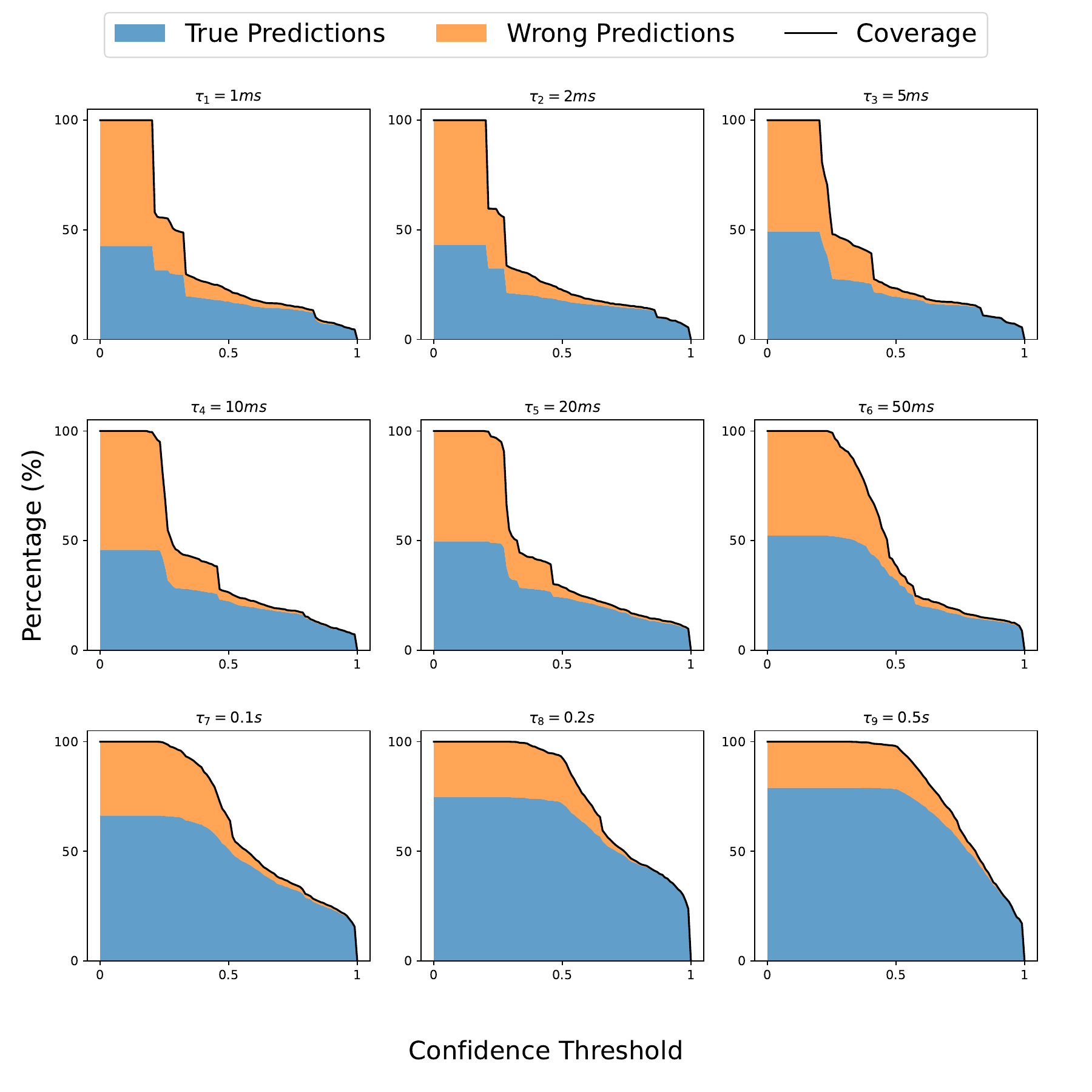}
    \caption{Graphs showing \emph{coverage} and percentage of true and false predictions as a function of the \emph{confidence} threshold, for classifiers of different exit times.}
    \label{fig:confidence}
\end{figure}

The results for a single classification task (namely, VPN Services with 5 bins) are shown in Figure \ref{fig:confidence}.
Some significant insights emerge from these results. First, even with short exit times, many samples have been classified with high confidence, thus validating the underlying principle of our method: \emph{many flows can be classified at an early stage}, thereby reducing the average classification time and saving memory resources.
Second, we observe an improvement in model coverage and accuracy over time, indicating that subsequent classifiers generally classify with higher accuracy and cover a broader range of flows.
Yet, some flows are correctly classified in earlier classifiers but are misclassified in subsequent ones. This indicates that there are timely features that become hard to recognize after a longer period.

Following this experiment, we proceeded to evaluate the \emph{EC} setup, where we determined the initial classification threshold $\beta$ based on the accuracy of a baseline model that classifies all flows after the maximal time $\tau_{max}$.\footnote{For clarity, in this section on \emph{EC}, we trained a single model using an $80\%{-}20\%$ random train-test split, omitting $k$-fold validation. This choice ensures comparability with uniform models employing 5-fold cross-validation, while also maintaining consistency with a single model and a single $\beta$ value.} 

\begin{figure}[tbp]
    \centering        
    \includegraphics[width=0.7\linewidth]{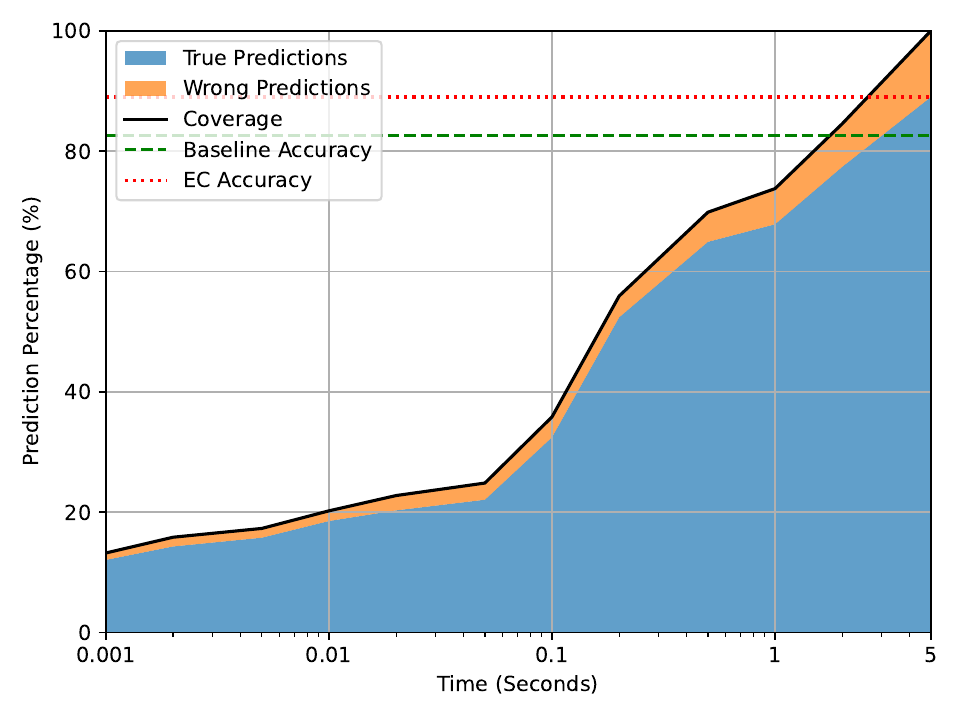}
    \caption{The percentage of flows classified up to each period (the \emph{coverage}) in the \emph{EC} setup.
    We can also see the amount of flows classified correctly and incorrectly. The final \emph{EC} model's accuracy is marked in red, while the predetermined threshold $\beta$ is marked in green.}
    \label{fig:ec}
\end{figure}

For the same classification task as shown in Figure \ref{fig:confidence} (VPN Services with 5 uniform bins), the \emph{EC} model results are shown in Figure \ref{fig:ec}. 

\begin{figure}[tbp]
    \centering
    \subfloat[Average Exit Time\label{fig:dist_size_vs_exit_time}]{    \includegraphics[width=0.6\linewidth]{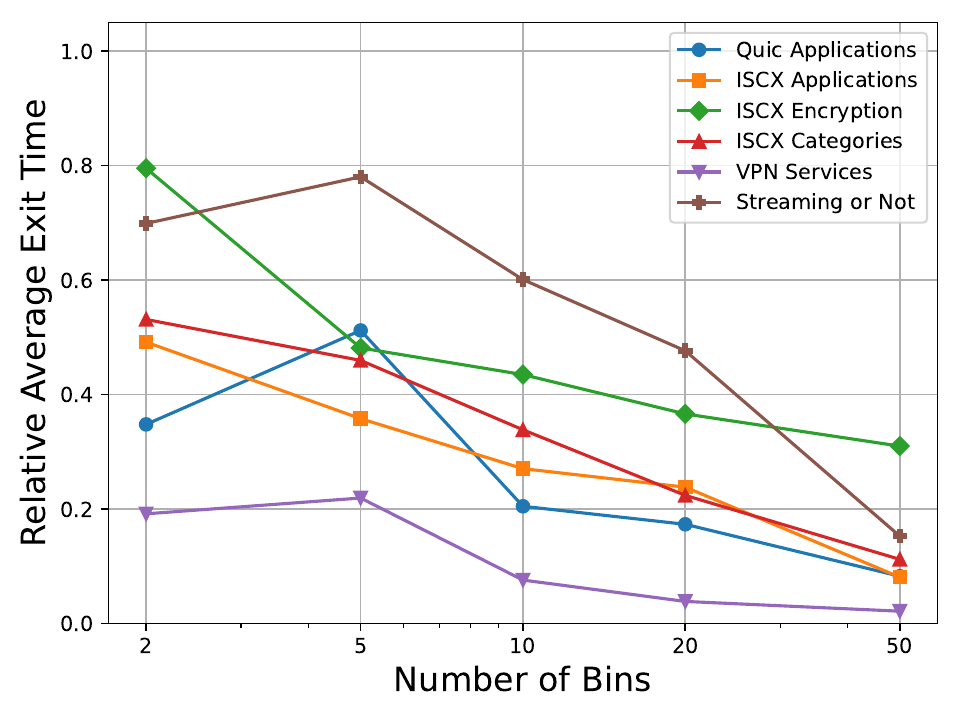}}
\qquad \qquad \qquad
    \subfloat[Accuracy\label{fig:dist_size_vs_accuracy}]{    \includegraphics[width=0.6\linewidth]{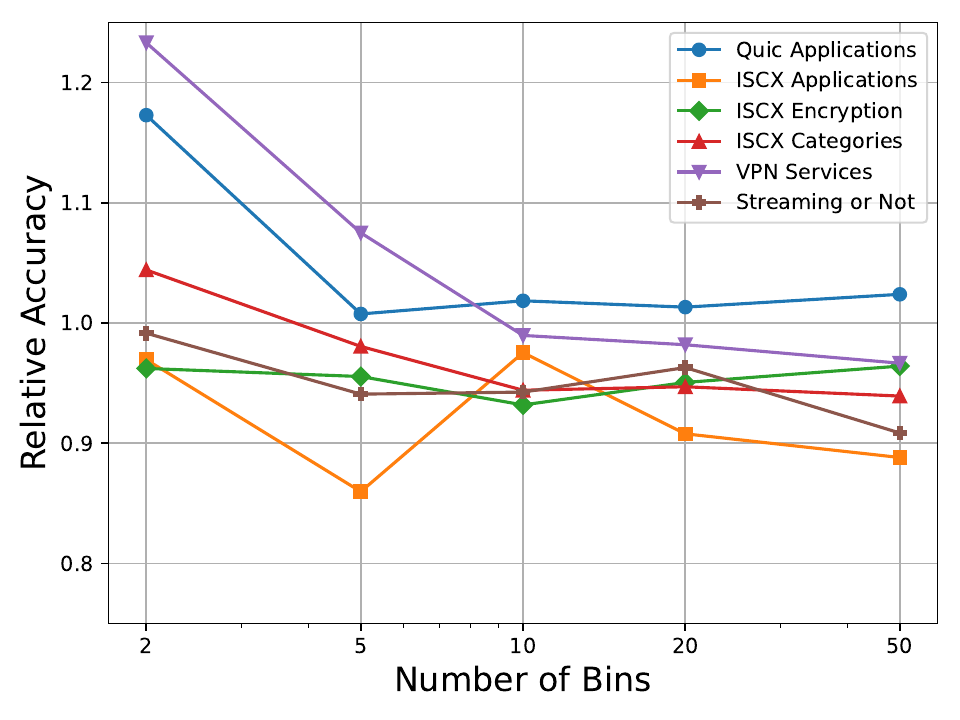}}
    
    \caption{The accuracy and the average exit time of \emph{EC} models with different numbers of bins, across all classification tasks. Results are relative to the maximal time period $\tau_{max}$ and the corresponding classifier $f_{max}$ of each specific classification task and number of bins.}

    \label{fig:dist_size_vs}
\end{figure}

We can see two phenomena. First, the average classification time for flows stands at $1.01$ seconds, in contrast to the baseline model's classification time of $5$ seconds, representing a notable improvement of approximately $80\%$ in classification speed.

Moreover, the final accuracy of the \emph{EC} model surpasses that of the baseline model. This means that not only we can save classification time, but \emph{EC} can potentially improve the model's accuracy. As previously discussed, this improvement could stem from timely features identified by earlier models, which may go unnoticed by subsequent ones.

Figure \ref{fig:dist_size_vs} compares models utilizing \emph{EC} for different tasks, where each value is normalized against a uniform binning model with the appropriate number of bins. 
These figures highlight the inherent trade-offs in employing $EC$: While larger representations facilitate faster classification, there is a slight compromise in accuracy compared to baseline models. On the other hand, smaller representations have the potential to surpass the performance of baseline models, as explained above. 

The rationale behind these results lies in the fact that larger representations encapsulate finer details about the flows, thereby boosting the confidence of earlier models. Nevertheless, the notable enhancement in reducing classification time comes with the potential of misclassification at these early stages.

\begin{figure}[tbp]
    \centering        
    \includegraphics[width=0.7\linewidth]{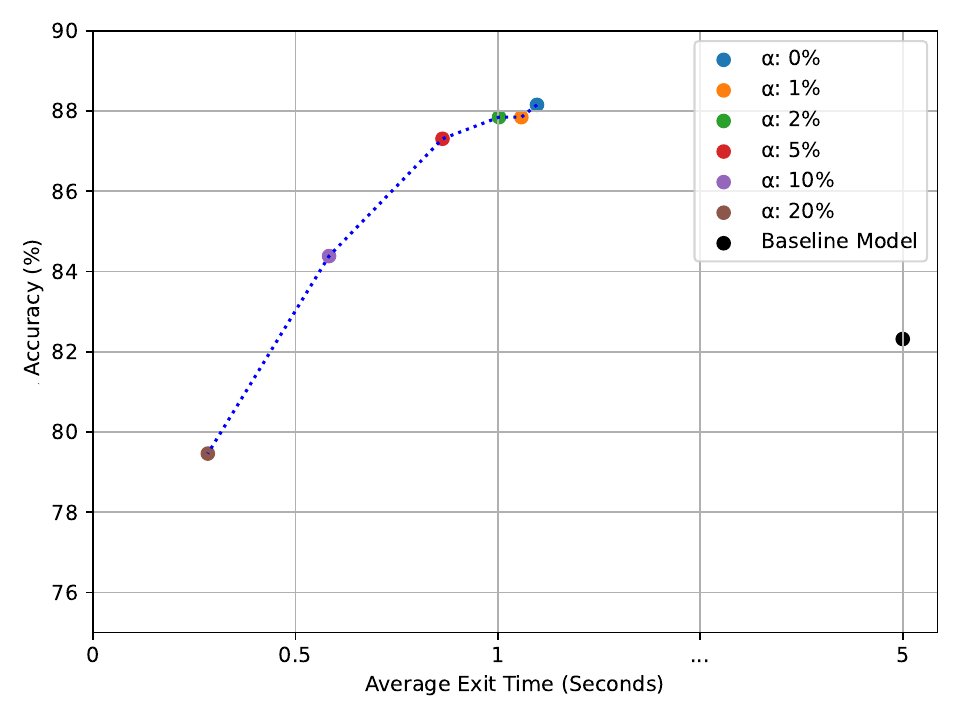}
    \caption{The accuracies and the average exit times of \emph{EC} models with different thresholds, for the VPN Services classification task with 5 bins. $\beta$ is set to be the accuracy of the baseline model, and the threshold for each model is $\beta - \alpha$.
    Notice that the graph's x-axis is not linear and shows the baseline model that classifies all flows after $\tau_{max}=5s$.
    }

    \label{fig:ee_alpha_comp}
\end{figure}

We also compare the results of the \emph{EC} setup with a lower classification threshold ($\beta-\alpha$). The results of this comparison for one classification task are shown in Figure \ref{fig:ee_alpha_comp}.

By opting for a lower threshold, a greater number of flows undergo early-stage classification, thereby reducing the average classification time.
However, while the classification time is reduced, there is a notable decrease in accuracy.
This tradeoff is controlled by selecting the threshold parameter, giving the flexibility to fine-tune this threshold according to specific requirements and constraints. 
In this experiment, it is evident that up to an $\alpha$ value of $5\%$, the decline in accuracy is negligible (lower than $1\%$), while there is a reduction of $25\%$ in collection time.

\begin{figure}[tbp]
    \centering
    \subfloat[VPN Services - 5 bins\label{fig:VPN_Services_5_log_alphas}]{\includegraphics[width=0.65\linewidth]{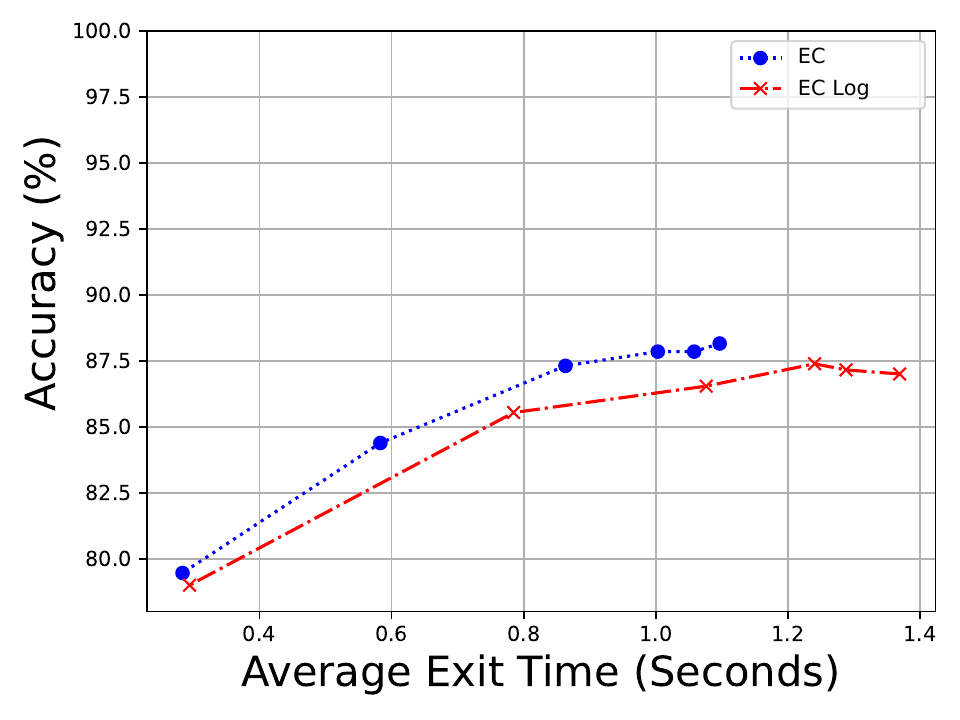}}
    \qquad\qquad\qquad
    \subfloat[VPN Services  - 10 bins\label{fig:VPN_Services_10_log_alphas}]{\includegraphics[width=0.65\linewidth]{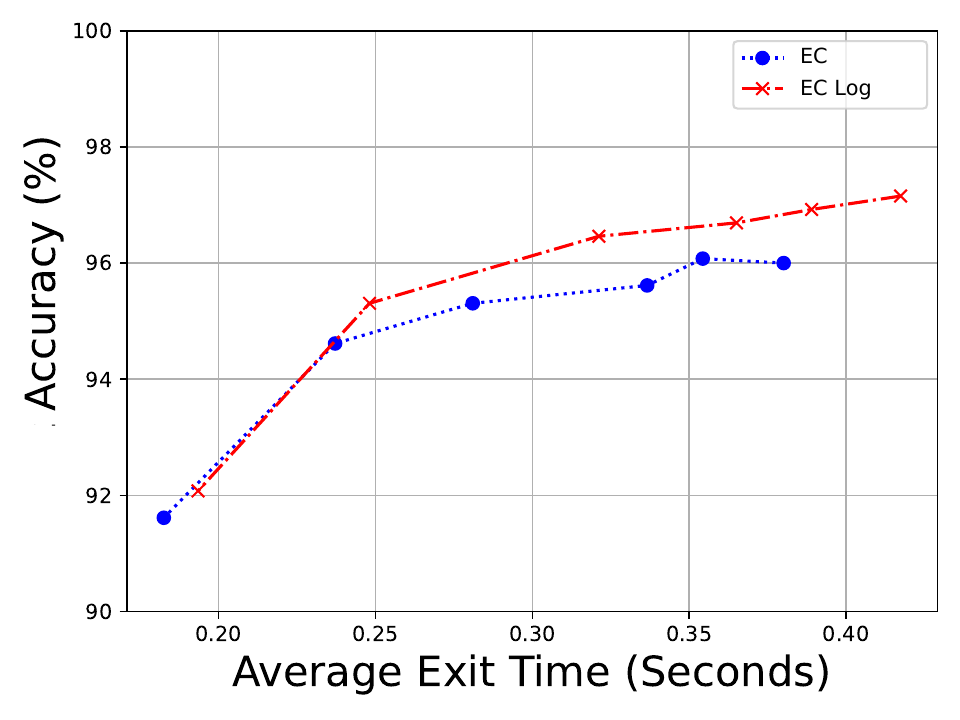}}
    \\
    \subfloat[VPN Services  - 20 bins\label{fig:VPN_Services_20_log_alphas}]{\includegraphics[width=0.65\linewidth]{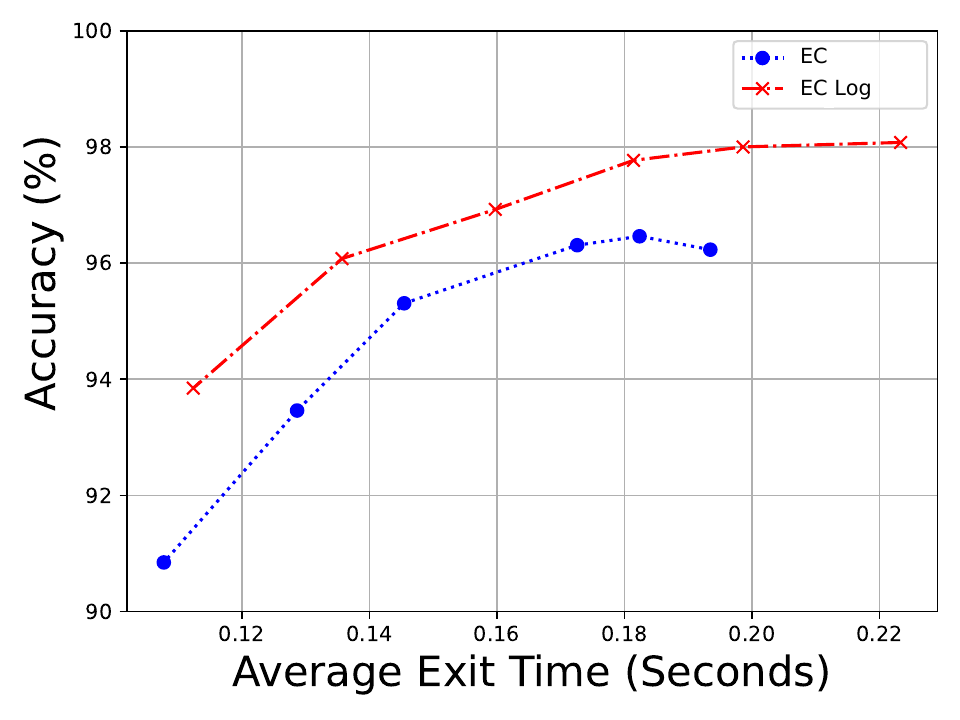}}
     \qquad\qquad\qquad
   \subfloat[VPN Services  - 50 bins\label{fig:VPN_Services_50_log_alphas}]{\includegraphics[width=0.65\linewidth]{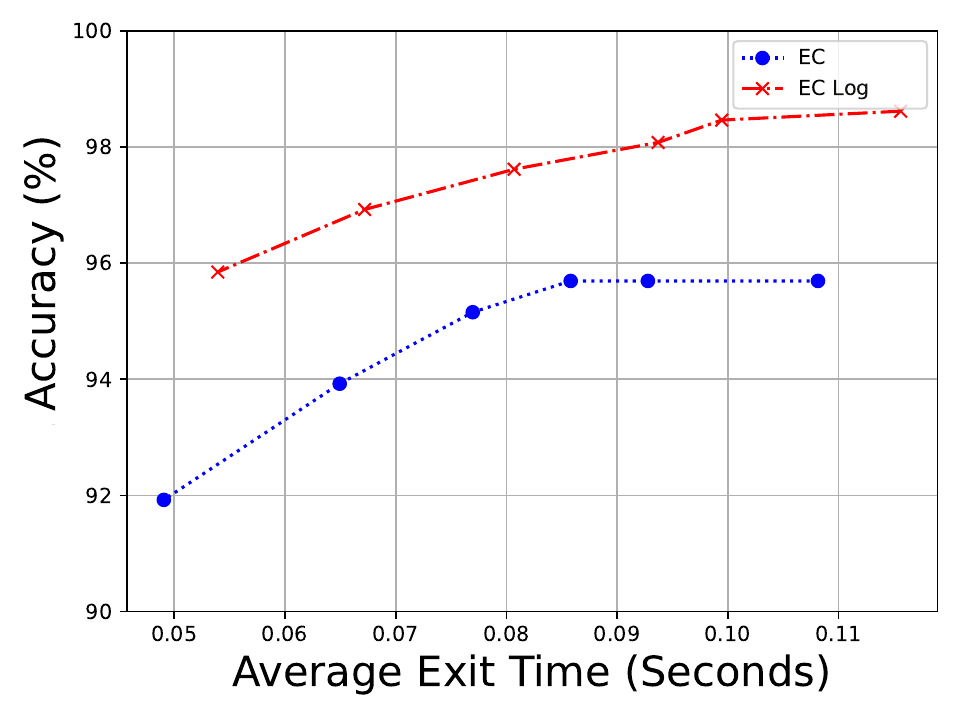}}
    \caption{Accuracy as a function of the average classification time, created using different $\alpha$ values (as in Figure \ref{fig:ee_alpha_comp}). Each graph compares the uniform and the logarithmic methods for creating the bins for the arrival-times dimension for a given number of bins.}
    \label{fig:log_vs_uniform}
\end{figure}

\subsection{Early Classification with Hyperparameter Optimization (ECHO)}
\label{sec:res:early_non_uniform}
Having noted the effectiveness of \emph{HO} in enhancing model accuracy (Section \ref{sec:res:non_uniform}), we aim to extend these representations to \emph{EC} setups as well. 

First, we will compare two methods for the bins of the arrival times distribution, the simple \emph{EC} with uniform time and size bins, and the \emph{EC Log} method with logarithmic time bins. The full description of those methods and how they are updated to allow real-time classification without memory overhead is in Section \ref{sec:early:non_uniform}.

The comparison results for different numbers of bins and different $\alpha$ values are shown in Figure \ref{fig:log_vs_uniform}.
The results for the two methods are similar. 
Yet, a noteworthy trend is that with smaller representations, the performance of \emph{EC Log} is inferior. Conversely, for larger representations, \emph{EC Log} tends to exhibit higher accuracy, surpassing the accuracy of the uniform \emph{EC}. 
We observe that while introducing non-uniformity in the time dimension enhances accuracy, its impact is relatively minor compared to the non-uniformity seen in the packet-sizes dimension, as depicted earlier in Figure \ref{fig:sizes_times_bins}.

In the following experiment, we evaluate the creation of \emph{EC} models with non-uniform bins in the packet size dimension, namely \emph{ECHO}. As explained in Section \ref{sec:early:non_uniform}, this process does not require any additional memory or computational power. 
For the \emph{ECHO} models in this comparison, $\beta$ is selected as the accuracy of a baseline uniform model, and $\alpha$ is $5\%$ based on previous observations (Figures \ref{fig:ee_alpha_comp} and \ref{fig:log_vs_uniform}). The selected $\alpha$ value strikes a balance between average exit time and accuracy, with similar trends noted for other values of $\alpha$.

\begin{figure}[tbp]
    \centering        
    \subfloat[Accuracy \label{fig:early_non_uniform_accuracy}]{    \includegraphics[width=0.65\linewidth]{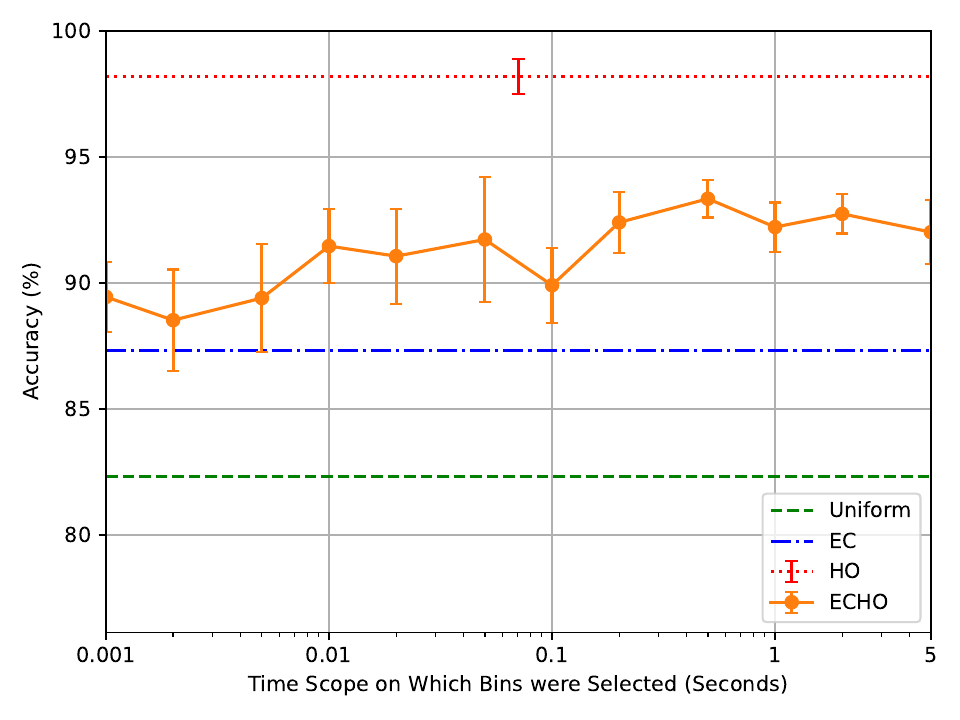}}
    \qquad\qquad\qquad
    \subfloat[Average exit time \label{fig:early_non_uniform_time}]{    \includegraphics[width=0.65\linewidth]{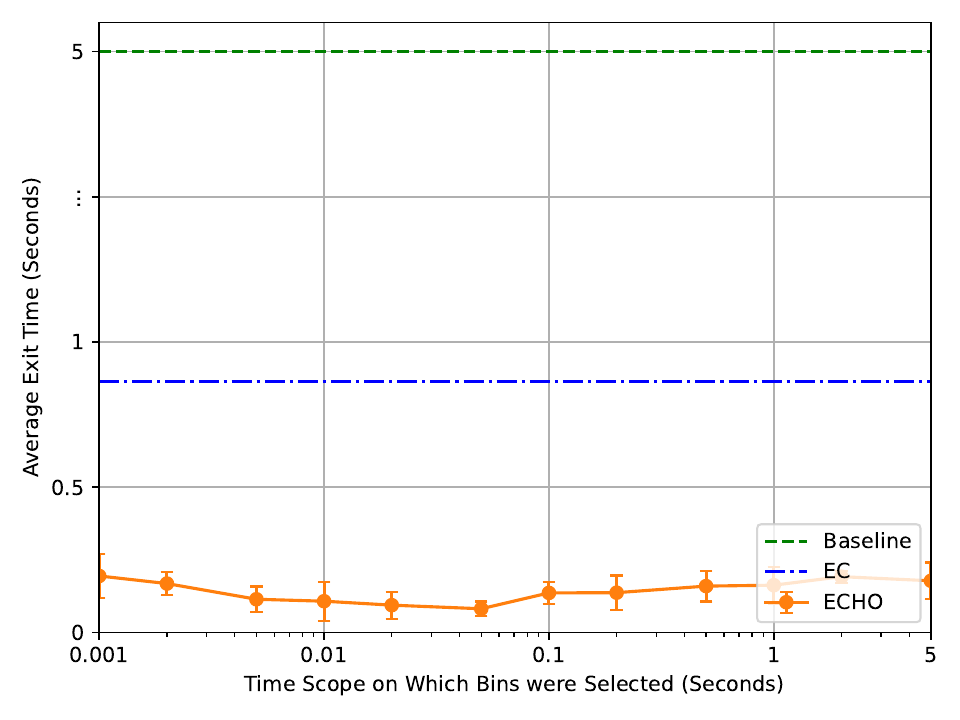}}

    \caption{Different bin selection methods for creating \emph{ECHO} models. Uniform and \emph{HO} models classify all flows after 5 seconds, while the \emph{EC} and \emph{ECHO} models perform early classification. \emph{ECHO} models have optimized packet size bins based on different exit times. The error bars in the Bayesian results show the standard deviation of the results of the \emph{NCV} process (see Section \ref{sec:non-uniform}). Notice that the time-graph's y-axis is not linear and shows the baseline models (Uniform or ECHO), classifying all flows after $\tau_{max}=5$. 
    }\label{fig:early_non_uniform}
\end{figure}

Figure \ref{fig:early_non_uniform} compares different selections of bins for the \emph{EC} model, where each selection is based on an \emph{HO} optimization over a classifier with the corresponding time scope $f_1,\ldots,f_{max}$. 
Figure \ref{fig:early_non_uniform_accuracy} compares the accuracy of the \emph{ECHO} setups, it is noticable that
\emph{ECHO} significantly improves the model's accuracy compared to the uniform \emph{EC} model. However, the accuracy results are still lower than an \emph{HO} setup (with non-uniform packet size bins) that classifies all flows after $\tau_{max}$.

Moreover, even though no specific classifier has a clear advantage in average accuracy, there is a smaller standard deviation for subsequent classifiers, implying that the binnings of these classifiers are more reliable and consistent, and therefore, are preferable.

In Figure \ref{fig:early_non_uniform_time}, we compare the same classifiers in terms of average exit time. Here, for all binnings, the \emph{ECHO} models show significantly faster average classification times (with no advantage to any specific selection). Recall that the threshold value for the \emph{EC} models is based on a baseline uniform model, and as shown earlier, \emph{HO} significantly improves accuracy, which is then reflected as well in higher confidence in classification. Thus enabling faster classification but with a lower accuracy than a full \emph{HO} model.

% maybe packet size vs accuracy \& time

%% file: chapters/conclusions/conclusions.tex
\section{Conclusions}
\label{sec:conclusions}
This paper highlights the fact that
many traffic classification techniques have considerable room for improvement. The techniques applied in the \emph{ECHO} approach can seamlessly be combined with many existing traffic classification methods, and reduce both their memory footprint and their collection time without compromising any aspect of the classification process. 

We show that by employing hyperparameter optimization methods to create non-uniform flow representations, we can significantly reduce the memory overhead of representations without compromising model accuracy. Furthermore, it boosts the model's throughput as smaller and more efficient classifiers are used. 

We also demonstrate that building a setup that includes multiple exit times with multiple classifiers, can significantly reduce the average required time for classification with no additional memory requirements. 

In this paper, we utilize the TPE algorithm to optimize the variable-sized bins. Although this process is conducted offline, in future work, we aim to investigate other methods with improved time efficiency. We also wish to explore whether other flow representations could be optimized in terms of memory usage.

In addition, we aim to delve into different methods to create the setup for early traffic classification, and different methods for assessing the models' confidence as part of the early classification process.

\subsection{Reproducability} Our code, as well as the processed datasets, will be publicly available to the community upon completion of the double-blind review process of our paper.

\subsection{Acknowledgments}
Part of this work has been supported by Israel
Innovation Authority within the ENTM consortium.

%% file: chapters/methods/datasets_appendix.tex
\section{Datasets}
\label{sec:datasets}
This section provides an overview of the three publicly available datasets used for training and testing our models, and the different classification tasks for each dataset.

\paragraph{Quic traffic dataset~\cite{rezaei2020achieve}.}
The Quic traffic dataset was recorded at UC Davis and consists of the traffic of 5 Google services (Google Drive, YouTube, Google Docs, Google Search,
and Google Music) under several operating systems (including different Windows and Ubuntu distributions). 
The flow counts are shown in Table~\ref{tab:quic flows}.

For this dataset, we consider one classification task, which we call \emph{Quic Applications}, of classifying flows to one of the different applications.
\paragraph{ISCX combined traffic dataset.}
This dataset consists of 2 network captures produced by the University of New Brunswick~\cite{inproceedings, HabibiLashkari2017CharacterizationOT}. %, as well as a small packet capture collected by Tel Aviv University~\cite{shapira2019flowpic} - in the end this was not used. 
The captures were filtered to include only the traffic related to the actual label of the flow. Classes with an insufficient number of samples (e.g., chat) were discarded. The processed dataset consists of 4 traffic categories (VoIP, video, file transfer, and browsing) with different applications for each category. 

\begin{table}[bt]
\caption{The number of flows per application in the Quic traffic dataset.}
\label{tab:quic flows}
\centering{
\small{
\begin{tabular}{|l|c|}
\toprule
\textbf{Application} & \textbf{Number of Flows} \\
\midrule
    \textbf{Google Search} & 1915 \\

    \textbf{Google Drive} & 1636 \\

    \textbf{Google Doc} & 1221 \\

    \textbf{Youtube} & 1077 \\

    \textbf{Google Music} & 592 \\
    \bottomrule
\end{tabular}
}}
\end{table}

For these categories, we
have 3 encryption techniques: non-VPN, VPN (for all classes except browsing), and TOR.
As proposed in the original papers, the dataset is sampled using a 15-second non-overlapping time window and every sample is issued as a different flow.
Table~\ref{tab:iscx flow} shows the number of flows for each category and each type of encryption, while Table~\ref{tab:iscx apps} delves into specific applications within the video and VoIP categories.
\begin{table}[tb]
\caption{The number of flows per category and encryption technique in the ISCX combined traffic dataset.}
\centering{
\small{
\label{tab:iscx flow}
\begin{tabular}{|c|c|c|c|}
\toprule
 \textbf{Category}& \textbf{Non-VPN} & \textbf{VPN} & \textbf{Tor} \\
\midrule
\textbf{Voip}
&  4397&  1543&  1761\\
\textbf{Video}
&  1180&  {301}&  {438}\\
\textbf{File Transfer}
&  {954}&  {303}&  {546}\\
\textbf{Browsing}
 & {757}& - & {448}\\
 \bottomrule
\end{tabular}
}}
\end{table}
\begin{table}[tb]
\caption{The number of flows per application in the ISCX traffic dataset (Non-VPN).}
\label{tab:iscx apps}
\centering{
\small{
\begin{tabular}%{|p{40px}|p{40px}|p{40px}|p{40px}|p{40px}|p{40px}|p{40px}|p{40px}|p{40px}|p{40px}|}
{|l|l|c|}
\toprule
\textbf{Category} & \textbf{Application} & \textbf{Number of flows} \\
\midrule
\multirow{4}{*}{\textbf{VoIP}} 
&\textbf{Facebook} & 1997\\
&\textbf{Hangout} & 2441\\
&\textbf{Skype} & 1823\\ &\textbf{Voipbuster}& 1032\\ 
\midrule
\multirow{6}{*}{\textbf{Video}} 
& \textbf{Vimeo} & 565 \\ &\textbf{YouTube} & 534\\ &\textbf{Hangouts} & 227\\ &\textbf{Skype} & 210\\ &\textbf{Netflix} & 207\\ &\textbf{Facebook} & 176\\
\bottomrule
\end{tabular}
}}
\end{table}

For this dataset, we consider three classification tasks: (i)  \emph{ISCX Applications} in which we classify the different applications in Table~\ref{tab:iscx apps}; (ii) \emph{ISCX Encryption}, in which we classify the different encryption protocols (namely, VPN, Non-VPN, or TOR); and (iii) \emph{ISCX Categories}, in which we obtain the category (namely, video, VoIP, file transfer, or browsing) for the flow.
\paragraph{VPN services dataset~\cite{NAAS2023108945}.}
This dataset consists of packet captures produced and processed by the University of South Bohemia. The processed dataset contains 5 traffic categories (email, video conferencing, SSH, streaming, and non-streaming) using different applications for the streaming and non-streaming categories (the specific applications are not available in the dataset). Traffic from all categories is recorded over VPN and non-VPN connections, and using 7 different VPN settings (L2TP, L2TP IPsec, OpenVPN, PPTP, SSTP, WireGuard, and non-VPN).

Since the dataset contains a low volume of email, video conferencing, and SSH traffic, we only use the streaming and non-streaming categories. We also exclude the traffic recorded over PPTP VPN as it contains significantly fewer samples than the other VPN services.
The number of flows in each category for the filtered dataset is shown in Table \ref{tab: vpn flows}.
\begin{table}[tb]
\caption{The number of flow by category and VPN service in the VPN service dataset~\cite{NAAS2023108945}.}
\label{tab: vpn flows}
\centering{
\small{
  \begin{tabular}{|l|c|c|}
    \toprule
    {\textbf{VPN Protocol}} & \textbf{non-streaming} & \textbf{streaming} \\
    \midrule
    \textbf{L2TP} & 1003 & 94 \\
    \textbf{L2TP IPsec} & 1003 & 93 \\
    \textbf{Non VPN} & 3277 & 440 \\
    \textbf{OpenVPN} & 999 & 96 \\
    \textbf{SSTP} & 986 & 97 \\
    \textbf{WireGuard} & 1003 & 99 \\
    \bottomrule
  \end{tabular}
  }}
\end{table}
For this dataset, we consider two classification tasks. First, \emph{VPN Services}, is the task of classifying the specific VPN in use. Secondly, \emph{Streaming or Not} is the binary task of categorizing the flows of the dataset into streaming/non-streaming categories.

\subsection{Preprocessing}

%or the training and evaluation of our models, we use k-fold cross-validation\cite{Kohavi1995ASO} with $k=5$, the dataset is divided into 5 equal-sized subsets or "folds" and we create 5 versions of our model, with each fold serving as a testing set once, while the remaining k-1 folds are used for training. We collect the mean and the standard deviation of the accuracy results.
% sliding window in older versions: sample each flow using a sliding time window (to increase the number of samples) and 
We filter the samples using three parameters: minimal number of packets, minimal volume (in bytes), and minimal duration. 
For every dataset, we have selected these parameters with regard to similar experiments done in the literature and to our observations of the traffic patterns. 
The selected parameters for each dataset are shown in Table \ref{tab: filtering params}.
\begin{table}[tb]
    \caption{Preprocessing parameters}
    \label{tab: filtering params}
    \centering{
\small{
\begin{tabular}{|l|c|c|c|c|}
    \toprule
     & \textbf{Timeout $\tau$} & \textbf{Min. Packets} & \textbf{Min. Vol.} & \textbf{Min. Dur.} \\
    \midrule
    \textbf{QUIC} & 1s & 10 & 0 & 0.5s \\
    \textbf{VPN} & 5s & 100 & 10KB & 3s \\
    \textbf{ISCX} & 15s & 100 & 10KB & 5s \\
    \bottomrule
  \end{tabular}
}}
\end{table}
Additionally, as some of our datasets are imbalanced, we have employed random undersampling~\cite{10.1145/1007730.1007735} to balance the data. This balancing procedure was applied to both the training and test sets to ensure equal representation of each class.

%% file: chapters/results/models.tex
\section{Representations and Models Comparison}
\label{sec:res:models}
In this section, we thoroughly compare different representations with various sizes (detailed below), employing 4 different classifiers, Random Forest (RF)~\cite{598994}, 
Logistic Regression (LR)~\cite{doi:10.1073/pnas.6.6.275}, 
Support Vector Machine (SVM)~\cite{Cortes1995SupportVectorN}, and Convolutional Neural Networks (CNNs, denoted here by DEEP)~\cite{Fukushima1980NeocognitronAS}. 
Although \emph{ECHO} can work with any distributional representation and classifier, this comparison justify our selection of representation and model in Section~\ref{sec:res}. 
%In this Section, we also justify our selection of model representation and classifier.

 Recall that we denote the selected representation type as $r$, where $r_q$ is the calculated representation for the flow with identifier $q$. 
 We compare the following four common types of flow representations.
% \label{sec:representations}
\begin{description}[leftmargin=0cm,labelindent=0cm,parsep=0.5em]
    \item[\emph{Distribution vectors},] denoted by $dist(N)$.
    This representation consists of 4 distribution vectors of size $N$, containing, for each flow direction, counters of the arrival times and packet sizes, where each value is mapped to one of the $N$ bins, as suggested by~\cite{engelberg2021classification}.
    $dist(N)$ works under the classifiers $DEEP$ 1D-CNN, $RF$, $LR$, and $SVM$.
    The size of $dist(N)$ is $4N$ bytes as we store 4 vectors of size $N$, using 8-bit integers.
    \item[\emph{Time series},] denoted by $ts(N)$. Time series of arrival time, packet size, and packet direction of $N$ packets, as proposed by \cite{10.1145/1129582.1129589,9585567}. $ts(N)$ works under the classifiers $DEEP$ 1D-CNN, $RF$, $LR$, and $SVM$.
    The size of $ts(N)$ is approximately $6N$ bytes as we store the arrival time (a 32-bit float), packet size (11 bits for sizes up to 1500), and direction (a boolean value) of $N$ packets.
    \item[\emph{FlowPic},] denoted by $fp(N\times N)$. The FlowPic representation is a two-dimensional distribution matrix of size $N\times N$ of arrival times and packet sizes, as proposed by~\cite{10.1145/3517745.3561436, shapira2019flowpic, engelberg2021classification, Finamore2023ReplicationCL}.
    As in $dist(N)$, each packet size and arrival time is mapped to one of $N$ bins. Then, element ($i$,$j$) in the matrix holds the number of packets whose arrival times are mapped to (arrival time) bin $i$, and packet sizes are mapped to (packet size) bin $j$.  
    
    $fp(N\times N)$ can be used under the classifiers $DEEP$ 2D-CNN, $RF$, $LR$, and $SVM$.
    The size of $fp(N\times N)$ is $2\cdot N^2$ bytes as we store a two-dimensional vector of size $N$ for each traffic side, using 8-bit integers.
    \item[\emph{Statistical Features},] denoted by $sts$. Statistical representation, containing 33 basic statistical features, as proposed by~\cite{article,inproceedings,HabibiLashkari2017CharacterizationOT,inproceedings_n} and others. 
    The features are the basic minimum, maximum, mean, median, and standard deviation of the packet sizes and packet relative arrival times, for all packets and each traffic direction independently, as well as the number of packets for each direction. 
    This representation is only used with simpler ML classifiers: $RF$, $LR$, and $SVM$. 
    The size of $sts$ is $132$ bytes if all statistics are stored as 32-bit float values (this can be reduced by setting specific types to each value)\footnote{Although statistics like median cannot be evaluated without saving the entire time-series, we assume that these parameters can be estimated accurately, e.g.~\cite{Roughan2004ClassofserviceMF}.}. 
    %\cite{BLUM1973448}.}.
\end{description}

\subsection{Flow Representations}
\label{sec:background:representations}

The results for the different classification tasks are shown in Tables 
% \ref{tab:tab:Quic_Apps_table}, \ref{tab:tab:ISCX_Apps_table}, \ref{tab:tab:ISCX_Encryption_table}, \ref{tab:tab:ISCX_Categories_table}, \ref{tab:tab:VPN_Services_table}, 
% and \ref{tab:tab:VPN_Strm_table}. 
\ref{tab:tab:Quic_Apps_table}--\ref{tab:tab:VPN_Strm_table}.
For brevity, we show only the accuracy and exclude the std results of the $k$-fold process. 
Additionally, we note that the neural network architecture of the $DEEP$ CNNs is not relevant for small $dist$ and $ts$ representations (as it includes pooling layers), as well as $sts$ representations (which have no temporal structure). 
\small{
\begin{table}[tb]
    \caption{QUIC Applications classification task.}
    \centering
    \begin{tabular}{|l|c|c|c|c|}
    \toprule
    & {\textbf{RF}} & {\textbf{SVM}} & {\textbf{LR}} & {\textbf{DEEP}} \\
    \midrule
    \textbf{dist(5)}& \textbf{94.9\%} & 82.7\% & 93.6\% & -\\
\textbf{dist(10)}& \textbf{95.9\%} & 82.3\% & 95.5\% & -\\
\textbf{dist(20)}& \textbf{95.9\%} & 82.2\% & 95.7\% & 94.9\%\\
\underline{\textbf{dist(50)}}& 95.7\% & 79.4\% & 94.4\% & \textbf{96.2\%}\\
\textbf{ts(5)}& \textbf{94.6\%} & 81.3\% & 81.2\% & -\\
\textbf{ts(10)}& \textbf{95.1\%} & 80.5\% & 79.3\% & -\\
\textbf{ts(20)}& \textbf{95.3\%} & 84.5\% & 79.7\% & 94.7\%\\
\textbf{ts(50)}& \textbf{95.8\%} & 85.5\% & 83.8\% & 93.3\%\\
\textbf{fp(8)}& 93.0\% & 78.7\% & \textbf{95.5\%} & 95.2\%\\
\textbf{fp(16)}& 91.8\% & 69.8\% & 94.7\% & \textbf{94.7\%}\\
\textbf{fp(32)}& 90.0\% & 66.9\% & 92.5\% & \textbf{95.5\%}\\
\textbf{sts()}& \textbf{95.0\%} & 82.9\% & 90.0\% & -\\
\bottomrule
    \end{tabular}
    \label{tab:tab:Quic_Apps_table}
    \end{table}
\begin{table}[tb]
    \caption{ISCX Applications classification task.}
    \centering
    \begin{tabular}{|l|c|c|c|c|}
    \toprule
    & {\textbf{RF}} & {\textbf{SVM}} & {\textbf{LR}} & {\textbf{DEEP}} \\
    \midrule
    \textbf{dist(5)}& \textbf{80.9\%} & 73.3\% & 76.4\% & -\\
\textbf{dist(10)}& 87.0\% & 79.9\% & \textbf{91.8\%} & -\\
\textbf{dist(20)}& 87.3\% & 82.8\% & 93.4\% & \textbf{95.9\%}\\
\underline{\textbf{dist(50)}}& 88.8\% & 91.7\% & 95.6\% & \textbf{97.8\%}\\
\textbf{ts(5)}& \textbf{67.0\%} & 49.5\% & 39.1\% & -\\
\textbf{ts(10)}& \textbf{68.5\%} & 56.2\% & 42.9\% & -\\
\textbf{ts(20)}& \textbf{69.6\%} & 60.1\% & 42.8\% & 53.5\%\\
\textbf{ts(50)}& 73.7\% & 61.1\% & 42.3\% & \textbf{73.9\%}\\
\textbf{fp(8)}& \textbf{80.4\%} & 72.3\% & 77.0\% & 78.0\%\\
\textbf{fp(16)}& 81.4\% & 74.9\% & 82.6\% & \textbf{91.4\%}\\
\textbf{fp(32)}& 83.0\% & 85.3\% & 84.2\% & \textbf{94.5\%}\\
\textbf{sts()}& 92.8\% & 69.3\% & \textbf{92.9\%} & -\\
\bottomrule
    \end{tabular}
    \label{tab:tab:ISCX_Apps_table}
    \end{table}
\begin{table}[tb]
    \caption{ISCX Encryption classification task.}
    \centering
    \begin{tabular}{|l|c|c|c|c|}
    \toprule
    & {\textbf{RF}} & {\textbf{SVM}} & {\textbf{LR}} & {\textbf{DEEP}} \\
    \midrule
    \textbf{dist(5)}& \textbf{92.3\%} & 83.6\% & 81.1\% & -\\
\textbf{dist(10)}& \textbf{92.7\%} & 85.5\% & 84.1\% & -\\
\textbf{dist(20)}& 92.8\% & 88.6\% & 90.2\% & \textbf{94.6\%}\\
\underline{\textbf{dist(50)}}& 90.8\% & 93.3\% & 91.6\% & \textbf{96.2\%}\\
\textbf{ts(5)}& \textbf{81.2\%} & 68.3\% & 60.8\% & -\\
\textbf{ts(10)}& \textbf{81.9\%} & 69.8\% & 59.0\% & -\\
\textbf{ts(20)}& \textbf{83.0\%} & 69.8\% & 58.0\% & 82.9\%\\
\textbf{ts(50)}& \textbf{84.6\%} & 69.8\% & 58.3\% & 84.5\%\\
\textbf{fp(8)}& \textbf{90.7\%} & 81.2\% & 84.5\% & 85.5\%\\
\textbf{fp(16)}& 92.1\% & 83.3\% & 88.4\% & \textbf{93.2\%}\\
\textbf{fp(32)}& 85.4\% & 81.4\% & 88.1\% & \textbf{92.7\%}\\
\textbf{sts()}& \textbf{95.2\%} & 78.4\% & 82.5\% & -\\
\bottomrule
    \end{tabular}
    \label{tab:tab:ISCX_Encryption_table}
    \end{table}
\begin{table}[tb]
    \caption{ISCX Categories classification task.}
    \centering
    \begin{tabular}{|l|c|c|c|c|}
    \toprule
    & {\textbf{RF}} & {\textbf{SVM}} & {\textbf{LR}} & {\textbf{DEEP}} \\
    \midrule
    \textbf{dist(5)}& \textbf{89.1\%} & 88.5\% & 82.6\% & -\\
\textbf{dist(10)}& \textbf{90.6\%} & 89.5\% & 89.4\% & -\\
\textbf{dist(20)}& 91.1\% & 90.6\% & \textbf{93.3\%} & 93.2\%\\
\textbf{dist(50)}& 89.4\% & 92.0\% & 91.5\% & \textbf{94.7\%}\\
\textbf{ts(5)}& \textbf{77.3\%} & 62.4\% & 55.3\% & -\\
\textbf{ts(10)}& \textbf{77.9\%} & 68.1\% & 61.5\% & -\\
\textbf{ts(20)}& 76.7\% & 72.8\% & 64.1\% & \textbf{78.3\%}\\
\textbf{ts(50)}& 76.5\% & 75.5\% & 63.9\% & \textbf{82.7\%}\\
\textbf{fp(8)}& 85.8\% & 89.0\% & 87.6\% & \textbf{92.2\%}\\
\textbf{fp(16)}& 83.1\% & 89.7\% & 88.7\% & \textbf{93.6\%}\\
\underline{\textbf{fp(32)}}& 80.7\% & 88.2\% & 88.5\% & \textbf{94.7\%}\\
\textbf{sts()}& \textbf{92.3\%} & 82.6\% & 86.5\% & -\\
\bottomrule
    \end{tabular}
    \label{tab:tab:ISCX_Categories_table}
    \end{table}
\begin{table}[tb]
    \caption{VPN Services classification task.}
    \centering
    \begin{tabular}{|l|c|c|c|c|}
    \toprule
    & {\textbf{RF}} & {\textbf{SVM}} & {\textbf{LR}} & {\textbf{DEEP}} \\
    \midrule
    \textbf{dist(5)}& \textbf{90.3\%} & 86.5\% & 82.3\% & -\\
\textbf{dist(10)}& 97.8\% & \textbf{98.7\%} & 97.1\% & -\\
\textbf{dist(20)}& \textbf{98.9\%} & 98.6\% & 98.4\% & 98.9\%\\
\underline{\textbf{dist(50)}}& \textbf{99.7\%} & 98.6\% & 99.0\% & 99.2\%\\
\textbf{ts(5)}& \textbf{92.2\%} & 58.3\% & 36.0\% & -\\
\textbf{ts(10)}& \textbf{91.6\%} & 63.2\% & 49.0\% & -\\
\textbf{ts(20)}& \textbf{92.4\%} & 73.5\% & 60.0\% & 91.6\%\\
\textbf{ts(50)}& \textbf{91.7\%} & 74.4\% & 62.0\% & 90.2\%\\
\textbf{fp(8)}& 96.2\% & 93.2\% & 95.9\% & \textbf{97.7\%}\\
\textbf{fp(16)}& 96.7\% & 97.5\% & 97.5\% & \textbf{99.2\%}\\
\textbf{fp(32)}& 95.4\% & 97.3\% & 97.2\% & \textbf{99.5\%}\\
\textbf{sts()}& \textbf{96.2\%} & 85.7\% & 93.4\% & -\\
\bottomrule
    \end{tabular}
    \label{tab:tab:VPN_Services_table}
    \end{table}
\begin{table}[tb]
    \caption{Streaming or Not classification task.}
    \centering
    \begin{tabular}{|l|c|c|c|c|}
    \toprule
    & {\textbf{RF}} & {\textbf{SVM}} & {\textbf{LR}} & {\textbf{DEEP}} \\
    \midrule
    \textbf{dist(5)}& \textbf{82.4\%} & 75.2\% & 69.8\% & -\\
\textbf{dist(10)}& \textbf{84.1\%} & 80.0\% & 79.1\% & -\\
\textbf{dist(20)}& \textbf{82.2\%} & 79.5\% & 79.4\% & 82.1\%\\
\textbf{dist(50)}& \textbf{80.4\%} & 77.8\% & 78.5\% & 79.9\%\\
\textbf{ts(5)}& \textbf{68.1\%} & 62.1\% & 59.3\% & -\\
\textbf{ts(10)}& \textbf{69.4\%} & 65.3\% & 64.3\% & -\\
\textbf{ts(20)}& \textbf{69.9\%} & 65.1\% & 64.4\% & 63.5\%\\
\textbf{ts(50)}& \textbf{69.5\%} & 62.6\% & 63.1\% & 65.1\%\\
\textbf{fp(8)}& 79.1\% & 76.9\% & 75.9\% & \textbf{80.7\%}\\
\textbf{fp(16)}& 79.1\% & 77.1\% & 78.5\% & \textbf{82.6\%}\\
\underline{\textbf{fp(32)}}& 76.3\% & 74.5\% & 76.2\% & \textbf{84.2\%}\\
\textbf{sts()}& \textbf{80.1\%} & 64.9\% & 71.7\% & -\\
\bottomrule
    \end{tabular}
    \label{tab:tab:VPN_Strm_table}
    \end{table}
}
We observe some common trends for most of the tasks and representations:

First, $dist$ representations achieve the highest accuracy for most classification tasks (with two anomalies favoring $fp$), with the $DEEP$ CNN classifier achieving the highest accuracy. It is apparent that the distributional features in $dist$ and $fp$ representations hold meaningful information for classification more than raw values as a time series or a collection of statistical values.

Moreover, larger representations tend to achieve higher accuracy, this is due to the granularity of the representations which allows identifying intrinsic behaviors unnoticeable in smaller or more coarse representations\footnote{This trend is limited for $RF$ classifiers, as they are configured to include the same number of weights (the number of decision trees and the number of nodes in each tree) for all representations. Therefore, the computational power of such classifiers might decay as the representations grow, as was shown back in 1968 by Hughes~\cite{Hughes1968OnTM}.}. 

%In Section \ref{sec:res:non_uniform}, we will explore creating smaller dist representations, but by employing Bayesian methods, we will adjust the representations so that we will capture the fine-grained details that are useful for classification.

Finally, small $LR$ classifiers tend to achieve a very low accuracy compared to other classifiers. However, this gap is significantly narrowed for larger representations, where $LR$ classifiers with larger representations achieve nearly equivalent results to those achieved by stronger $DEEP$ classifiers. 

%We aim to achieve accuracy close to that of stronger classifiers but with minimal memory requirements and computational power, so that such a classifier will perform well at high network rates.

\subsection{Models' Deployment Considerations}
\label{sec:res:throughput}
In this section, we discuss the real-time deployment consideration of our classifiers. Particularly, we examine the \emph{memory requirements} and the \emph{throughput} of different representations and classifiers.
First, we compare the memory required for storing all flow representations until classification. This is factored by the flow rate in the network, the collection time $\tau$, and by the representation size.
We note that the memory required for storing the parameters of most classifiers is negligible compared to the flow representations.

Table~\ref{tab:memory} shows an approximation of the total required memory for storing flow representations\footnote{Note that the memory required for $ts$ representations could potentially be lower than the maximum value indicated in the table, given that classification may occur once the necessary number of packets arrive, before the collection time $\tau$.}. We can see that some representations (like larger $dist$ and $fp$ representations) may be impractical for real-time use on edge devices in a large network.
\begin{table}[tb]
\caption{Memory requirements for representation types (for the ISCX classification tasks), with $\tau = 15s$ and a flow rate of 1 million flows per second.}
\centering
\small
\begin{tabular}{|l|c|c|c|}
\toprule & {\textbf{Repr. Size (B)}}& {\textbf{Approx. Required Memory (B)}}\\ \midrule\
\textbf{dist(5)}& 20 & 300.0M\\
\textbf{dist(10)}& 40 & 600.0M\\
\textbf{dist(20)}& 80 & 1.2G\\
\textbf{dist(50)}& 200 & 3.0G\\
\textbf{ts(5)}& 30 & 450.0M\\
\textbf{ts(10)}& 60 & 900.0M\\
\textbf{ts(20)}& 120 & 1.8G\\
\textbf{ts(50)}& 300 & 4.5G\\
\textbf{fp(8)}& 128 & 1.9G\\
\textbf{fp(16)}& 512 & 7.7G\\
\textbf{fp(32)}& 2048 & 30.7G\\
\textbf{sts()}& 132 & 2.0G\\
\bottomrule
\end{tabular}
\label{tab:memory}
\end{table}

The second objective we explore is the \emph{throughput} of the models. That is, the number of flows processed by a classifier within a specified timeframe, measured based on given computational capabilities.
Specifically, we measured the number of flows classified within a 60-second period, using batches of 1000 samples. This method allows classifiers to perform batch operations more efficiently than single-sample classification (e.g. matrix multiplications).

\begin{table}[tbh]
    \caption{Classification throughput (in flows per second, using batches of 1K representations) for the ISCX Applications classification task.}
    \centering
    \begin{tabular}{|l|c|c|c|c|}
    \toprule
    & {\textbf{RF}} & {\textbf{SVM}} & {\textbf{LR}} & {\textbf{DEEP}} \\
    \midrule
\textbf{dist(5)}& 96.3K & 15.4K & \textbf{7.7M} & -\\
\textbf{dist(10)}& 95.7K & 13.8K & \textbf{1.7M} & -\\
\textbf{dist(20)}& 93.7K & 11.6K & \textbf{1.4M} & 9.2K\\
\textbf{dist(50)}& 89.2K & 10.1K & \textbf{894.2K} & 8.7K\\
\textbf{ts(5)}& 95.5K & 9.5K & \textbf{7.4M} & -\\
\textbf{ts(10)}& 94.2K & 9.4K & \textbf{1.8M} & -\\
\textbf{ts(20)}& 91.0K & 8.2K & \textbf{1.4M} & 9.2K\\
\textbf{ts(50)}& 84.9K & 6.1K & \textbf{937.1K} & 8.8K\\
\textbf{sts()}& 94.2K & 10.9K & \textbf{1.8M} & -\\
\textbf{fp(8)}& 91.8K & 10.1K & \textbf{1.4M} & 9.0K\\
\textbf{fp(16)}& 73.5K & 3.5K & \textbf{551.9K} & 7.8K\\
\textbf{fp(32)}& 53.3K & 766.7 & \textbf{177.6K} & 6.1K\\
\bottomrule
    \end{tabular}
    \label{tab:ISCX_Apps_throughput}
    \end{table}

Table \ref{tab:ISCX_Apps_throughput} shows the \emph{throughput} of different classifiers for different representations. It only includes the \emph{throughput} calculated on models trained on the ISCX APPS classification task, however, as the required computational power merely changes across datasets or classification tasks, we observed similar results for the other tasks. 
The absolute numbers strongly depend on the deployment setting which in our case was an otherwise unloaded machine with an Intel core i7-1165G7 processor (@ 2.80GHz 1.69GHz) with 16GB RAM, using Python’s Keras
for $DEEP$ CNNs (with tensorflow CPU backend) and SKlearn for $RF$, $LR$, and $SVM$ classifiers.

We can see that despite the high accuracy achieved by $DEEP$ CNN classifiers, the classification throughput may be impractical if deployed in a large ISP network with millions of new flows per second. On the other hand, we can see that as expected, $LR$ classifiers outperform other classifiers in terms of throughput by an order of magnitude.

%% file: main.bbl

\begin{thebibliography}{60}

%%% the Web address.

\ifx \showCODEN    \undefined \def \showCODEN     #1{\unskip}     \fi
\ifx \showDOI      \undefined \def \showDOI       #1{#1}\fi
\ifx \showISBNx    \undefined \def \showISBNx     #1{\unskip}     \fi
\ifx \showISBNxiii \undefined \def \showISBNxiii  #1{\unskip}     \fi
\ifx \showISSN     \undefined \def \showISSN      #1{\unskip}     \fi
\ifx \showLCCN     \undefined \def \showLCCN      #1{\unskip}     \fi
\ifx \shownote     \undefined \def \shownote      #1{#1}          \fi
\ifx \showarticletitle \undefined \def \showarticletitle #1{#1}   \fi
\ifx \showURL      \undefined \def \showURL       {\relax}        \fi
% invisible to TeX
\providecommand\bibfield[2]{#2}
\providecommand\bibinfo[2]{#2}
\providecommand\natexlab[1]{#1}
\providecommand\showeprint[2][]{arXiv:#2}

\bibitem[Aceto et~al\mbox{.}(2019)]%
        {ACETO2019106944}
\bibfield{author}{\bibinfo{person}{Giuseppe Aceto}, \bibinfo{person}{Domenico Ciuonzo}, \bibinfo{person}{Antonio Montieri}, {and} \bibinfo{person}{Antonio Pescapè}.} \bibinfo{year}{2019}\natexlab{}.
\newblock \showarticletitle{{MIMETIC}: Mobile encrypted traffic classification using multimodal deep learning}.
\newblock \bibinfo{journal}{\emph{Computer Networks}}  \bibinfo{volume}{165} (\bibinfo{year}{2019}), \bibinfo{pages}{106944}.
\newblock
\showISSN{1389-1286}


\bibitem[Aceto et~al\mbox{.}(2010)]%
        {Aceto2010PortLoadTT}
\bibfield{author}{\bibinfo{person}{Giuseppe Aceto}, \bibinfo{person}{Alberto Dainotti}, \bibinfo{person}{Walter de Donato}, {and} \bibinfo{person}{Antonio Pescap{\'e}}.} \bibinfo{year}{2010}\natexlab{}.
\newblock \showarticletitle{PortLoad: Taking the Best of Two Worlds in Traffic Classification}.
\newblock \bibinfo{journal}{\emph{2010 INFOCOM IEEE Conference on Computer Communications Workshops}} (\bibinfo{year}{2010}), \bibinfo{pages}{1--5}.
\newblock


\bibitem[Akbari et~al\mbox{.}(2021)]%
        {10.1145/3447382}
\bibfield{author}{\bibinfo{person}{Iman Akbari}, \bibinfo{person}{Mohammad~A. Salahuddin}, \bibinfo{person}{Leni Ven}, \bibinfo{person}{Noura Limam}, \bibinfo{person}{Raouf Boutaba}, \bibinfo{person}{Bertrand Mathieu}, \bibinfo{person}{Stephanie Moteau}, {and} \bibinfo{person}{Stephane Tuffin}.} \bibinfo{year}{2021}\natexlab{}.
\newblock \showarticletitle{A Look Behind the Curtain: Traffic Classification in an Increasingly Encrypted Web}.
\newblock \bibinfo{journal}{\emph{Proc. ACM Meas. Anal. Comput. Syst.}} \bibinfo{volume}{5}, \bibinfo{number}{1}, Article \bibinfo{articleno}{04} (\bibinfo{date}{feb} \bibinfo{year}{2021}), \bibinfo{numpages}{26}~pages.
\newblock


\bibitem[Anderson and McGrew(2017)]%
        {inproceedings_n}
\bibfield{author}{\bibinfo{person}{Blake Anderson} {and} \bibinfo{person}{David McGrew}.} \bibinfo{year}{2017}\natexlab{}.
\newblock \showarticletitle{Machine Learning for Encrypted Malware Traffic Classification: Accounting for Noisy Labels and Non-Stationarity}. In \bibinfo{booktitle}{\emph{KDD'17}}. \bibinfo{pages}{1723--1732}.
\newblock
\showISBNx{978-1-4503-4887-4}
\urldef\tempurl%
\url{https://doi.org/10.1145/3097983.3098163}
\showDOI{\tempurl}


\bibitem[Anonymous(2023)]%
        {private}
\bibfield{author}{\bibinfo{person}{Anonymous}.} \bibinfo{year}{2023}\natexlab{}.
\newblock
\newblock
\newblock
\shownote{Private Communincation.}.


\bibitem[Barradas et~al\mbox{.}(2021)]%
        {Barradas2021FlowLensEE}
\bibfield{author}{\bibinfo{person}{Diogo Barradas}, \bibinfo{person}{Nuno Santos}, \bibinfo{person}{Lu{\'i}s Rodrigues}, \bibinfo{person}{Salvatore Signorello}, \bibinfo{person}{Fernando M.~V. Ramos}, {and} \bibinfo{person}{Andr{\'e} Madeira}.} \bibinfo{year}{2021}\natexlab{}.
\newblock \showarticletitle{{FlowLens}: Enabling Efficient Flow Classification for {ML}-based Network Security Applications}. In \bibinfo{booktitle}{\emph{NDSS'21}}.
\newblock


\bibitem[Batista et~al\mbox{.}(2004)]%
        {10.1145/1007730.1007735}
\bibfield{author}{\bibinfo{person}{Gustavo E. A. P.~A. Batista}, \bibinfo{person}{Ronaldo~C. Prati}, {and} \bibinfo{person}{Maria~Carolina Monard}.} \bibinfo{year}{2004}\natexlab{}.
\newblock \showarticletitle{A Study of the Behavior of Several Methods for Balancing Machine Learning Training Data}.
\newblock \bibinfo{journal}{\emph{SIGKDD Explor. Newsl.}} \bibinfo{volume}{6}, \bibinfo{number}{1} (\bibinfo{date}{jun} \bibinfo{year}{2004}), \bibinfo{pages}{20–29}.
\newblock
\showISSN{1931-0145}
\urldef\tempurl%
\url{https://doi.org/10.1145/1007730.1007735}
\showDOI{\tempurl}


\bibitem[Bergstra et~al\mbox{.}(2011)]%
        {NIPS2011_86e8f7ab}
\bibfield{author}{\bibinfo{person}{James Bergstra}, \bibinfo{person}{R\'{e}mi Bardenet}, \bibinfo{person}{Yoshua Bengio}, {and} \bibinfo{person}{Bal\'{a}zs K\'{e}gl}.} \bibinfo{year}{2011}\natexlab{}.
\newblock \showarticletitle{Algorithms for Hyper-Parameter Optimization}. In \bibinfo{booktitle}{\emph{Advances in Neural Information Processing Systems}}, Vol.~\bibinfo{volume}{24}.
\newblock


\bibitem[Bergstra et~al\mbox{.}(2013)]%
        {pmlr-v28-bergstra13}
\bibfield{author}{\bibinfo{person}{James Bergstra}, \bibinfo{person}{Daniel Yamins}, {and} \bibinfo{person}{David Cox}.} \bibinfo{year}{2013}\natexlab{}.
\newblock \showarticletitle{Making a Science of Model Search: Hyperparameter Optimization in Hundreds of Dimensions for Vision Architectures}. In \bibinfo{booktitle}{\emph{ICML'13}} \emph{(\bibinfo{series}{Proceedings of Machine Learning Research}, Vol.~\bibinfo{volume}{28})}. \bibinfo{publisher}{PMLR}, \bibinfo{pages}{115--123}.
\newblock


\bibitem[Bernaille et~al\mbox{.}(2006a)]%
        {10.1145/1129582.1129589}
\bibfield{author}{\bibinfo{person}{Laurent Bernaille}, \bibinfo{person}{Renata Teixeira}, \bibinfo{person}{Ismael Akodkenou}, \bibinfo{person}{Augustin Soule}, {and} \bibinfo{person}{Kave Salamatian}.} \bibinfo{year}{2006}\natexlab{a}.
\newblock \showarticletitle{Traffic Classification on the Fly}.
\newblock \bibinfo{journal}{\emph{SIGCOMM Comput. Commun. Rev.}} \bibinfo{volume}{36}, \bibinfo{number}{2} (\bibinfo{date}{apr} \bibinfo{year}{2006}), \bibinfo{pages}{23–26}.
\newblock
\showISSN{0146-4833}


\bibitem[Bernaille et~al\mbox{.}(2006b)]%
        {OnTheFly}
\bibfield{author}{\bibinfo{person}{Laurent Bernaille}, \bibinfo{person}{Renata Teixeira}, \bibinfo{person}{Ismael Akodkenou}, \bibinfo{person}{Augustin Soule}, {and} \bibinfo{person}{Kave Salamatian}.} \bibinfo{year}{2006}\natexlab{b}.
\newblock \showarticletitle{Traffic classification on the fly}.
\newblock \bibinfo{journal}{\emph{SIGCOMM Comput. Commun. Rev.}} \bibinfo{volume}{36}, \bibinfo{number}{2} (\bibinfo{date}{apr} \bibinfo{year}{2006}), \bibinfo{pages}{23–26}.
\newblock
\showISSN{0146-4833}
\urldef\tempurl%
\url{https://doi.org/10.1145/1129582.1129589}
\showDOI{\tempurl}


\bibitem[Bernaille et~al\mbox{.}(2006c)]%
        {Bernaille2006EarlyAI}
\bibfield{author}{\bibinfo{person}{Laurent Bernaille}, \bibinfo{person}{Renata~Cruz Teixeira}, {and} \bibinfo{person}{Kave Salamatian}.} \bibinfo{year}{2006}\natexlab{c}.
\newblock \showarticletitle{Early application identification}. In \bibinfo{booktitle}{\emph{Conference on Emerging Network Experiment and Technology}}.
\newblock
\urldef\tempurl%
\url{https://api.semanticscholar.org/CorpusID:496969}
\showURL{%
\tempurl}


\bibitem[Bronzino et~al\mbox{.}(2021)]%
        {10.1145/3491052}
\bibfield{author}{\bibinfo{person}{Francesco Bronzino}, \bibinfo{person}{Paul Schmitt}, \bibinfo{person}{Sara Ayoubi}, \bibinfo{person}{Hyojoon Kim}, \bibinfo{person}{Renata Teixeira}, {and} \bibinfo{person}{Nick Feamster}.} \bibinfo{year}{2021}\natexlab{}.
\newblock \showarticletitle{Traffic Refinery: Cost-Aware Data Representation for Machine Learning on Network Traffic}.
\newblock \bibinfo{journal}{\emph{Proc. ACM Meas. Anal. Comput. Syst.}} \bibinfo{volume}{5}, \bibinfo{number}{3}, Article \bibinfo{articleno}{40} (\bibinfo{date}{dec} \bibinfo{year}{2021}), \bibinfo{numpages}{24}~pages.
\newblock


\bibitem[Cawley and Talbot(2010)]%
        {10.5555/1756006.1859921}
\bibfield{author}{\bibinfo{person}{Gavin~C. Cawley} {and} \bibinfo{person}{Nicola~L.C. Talbot}.} \bibinfo{year}{2010}\natexlab{}.
\newblock \showarticletitle{On Over-Fitting in Model Selection and Subsequent Selection Bias in Performance Evaluation}.
\newblock \bibinfo{journal}{\emph{J. Mach. Learn. Res.}}  \bibinfo{volume}{11} (\bibinfo{date}{aug} \bibinfo{year}{2010}), \bibinfo{pages}{2079–2107}.
\newblock
\showISSN{1532-4435}


\bibitem[Cortes and Vapnik(1995)]%
        {Cortes1995SupportVectorN}
\bibfield{author}{\bibinfo{person}{Corinna Cortes} {and} \bibinfo{person}{Vladimir~Naumovich Vapnik}.} \bibinfo{year}{1995}\natexlab{}.
\newblock \showarticletitle{Support-Vector Networks}.
\newblock \bibinfo{journal}{\emph{Machine Learning}}  \bibinfo{volume}{20} (\bibinfo{year}{1995}), \bibinfo{pages}{273--297}.
\newblock


\bibitem[Draper-Gil et~al\mbox{.}(2016)]%
        {inproceedings}
\bibfield{author}{\bibinfo{person}{Gerard Draper-Gil}, \bibinfo{person}{Arash~Habibi Lashkari}, \bibinfo{person}{Mohammad Saiful~Islam Mamun}, {and} \bibinfo{person}{Ali~A Ghorbani}.} \bibinfo{year}{2016}\natexlab{}.
\newblock \showarticletitle{Characterization of encrypted and {VPN} traffic using time-related features}. In \bibinfo{booktitle}{\emph{ICISSP}}. \bibinfo{pages}{407--414}.
\newblock


\bibitem[Engelberg and Wool(2021)]%
        {engelberg2021classification}
\bibfield{author}{\bibinfo{person}{Aviv Engelberg} {and} \bibinfo{person}{Avishai Wool}.} \bibinfo{year}{2021}\natexlab{}.
\newblock \bibinfo{title}{Classification of Encrypted IoT Traffic Despite Padding and Shaping}.
\newblock
\newblock
\showeprint[arxiv]{2110.11188}~[cs.CR]
\urldef\tempurl%
\url{https://arxiv.org/abs/2110.11188}
\showURL{%
\tempurl}


\bibitem[Enghardt et~al\mbox{.}(2020)]%
        {rfc8922}
\bibfield{author}{\bibinfo{person}{Reese Enghardt}, \bibinfo{person}{Tommy Pauly}, \bibinfo{person}{Colin Perkins}, \bibinfo{person}{Kyle Rose}, {and} \bibinfo{person}{Christopher~A. Wood}.} \bibinfo{year}{2020}\natexlab{}.
\newblock \bibinfo{title}{{A Survey of the Interaction between Security Protocols and Transport Services}}.
\newblock \bibinfo{howpublished}{RFC 8922}.
\newblock


\bibitem[Ertam and Avcı(2017)]%
        {ERTAM2017135}
\bibfield{author}{\bibinfo{person}{Fatih Ertam} {and} \bibinfo{person}{Engin Avcı}.} \bibinfo{year}{2017}\natexlab{}.
\newblock \showarticletitle{A new approach for internet traffic classification: {GA-WK-ELM}}.
\newblock \bibinfo{journal}{\emph{Measurement}}  \bibinfo{volume}{95} (\bibinfo{year}{2017}), \bibinfo{pages}{135--142}.
\newblock
\showISSN{0263-2241}
\urldef\tempurl%
\url{https://doi.org/10.1016/j.measurement.2016.10.001}
\showDOI{\tempurl}


\bibitem[Fahad et~al\mbox{.}(2013)]%
        {FAHAD20132040}
\bibfield{author}{\bibinfo{person}{Adil Fahad}, \bibinfo{person}{Zahir Tari}, \bibinfo{person}{Ibrahim Khalil}, \bibinfo{person}{Ibrahim Habib}, {and} \bibinfo{person}{Hussein Alnuweiri}.} \bibinfo{year}{2013}\natexlab{}.
\newblock \showarticletitle{Toward an efficient and scalable feature selection approach for internet traffic classification}.
\newblock \bibinfo{journal}{\emph{Computer Networks}} \bibinfo{volume}{57}, \bibinfo{number}{9} (\bibinfo{year}{2013}), \bibinfo{pages}{2040--2057}.
\newblock
\showISSN{1389-1286}
\urldef\tempurl%
\url{https://doi.org/10.1016/j.comnet.2013.04.005}
\showDOI{\tempurl}


\bibitem[Finamore et~al\mbox{.}(2023)]%
        {Finamore2023ReplicationCL}
\bibfield{author}{\bibinfo{person}{Alessandro Finamore}, \bibinfo{person}{Chao Wang}, \bibinfo{person}{Jonatan Krolikowski}, \bibinfo{person}{Jos{\'e}~Manuel Navarro}, \bibinfo{person}{Fuxing Chen}, {and} \bibinfo{person}{Dario Rossi}.} \bibinfo{year}{2023}\natexlab{}.
\newblock \showarticletitle{Replication: Contrastive Learning and Data Augmentation in Traffic Classification Using a Flowpic Input Representation}.
\newblock \bibinfo{journal}{\emph{Proceedings of the 2023 ACM on Internet Measurement Conference}} (\bibinfo{year}{2023}).
\newblock
\urldef\tempurl%
\url{https://api.semanticscholar.org/CorpusID:262046478}
\showURL{%
\tempurl}


\bibitem[Finsterbusch et~al\mbox{.}(2014)]%
        {6644335}
\bibfield{author}{\bibinfo{person}{Michael Finsterbusch}, \bibinfo{person}{Chris Richter}, \bibinfo{person}{Eduardo Rocha}, \bibinfo{person}{Jean-Alexander Muller}, {and} \bibinfo{person}{Klaus Hanssgen}.} \bibinfo{year}{2014}\natexlab{}.
\newblock \showarticletitle{A Survey of Payload-Based Traffic Classification Approaches}.
\newblock \bibinfo{journal}{\emph{IEEE Communications Surveys \& Tutorials}} \bibinfo{volume}{16}, \bibinfo{number}{2} (\bibinfo{year}{2014}), \bibinfo{pages}{1135--1156}.
\newblock
\urldef\tempurl%
\url{https://doi.org/10.1109/SURV.2013.100613.00161}
\showDOI{\tempurl}


\bibitem[Fukushima(1980)]%
        {Fukushima1980NeocognitronAS}
\bibfield{author}{\bibinfo{person}{Kunihiko Fukushima}.} \bibinfo{year}{1980}\natexlab{}.
\newblock \showarticletitle{Neocognitron: A self-organizing neural network model for a mechanism of pattern recognition unaffected by shift in position}.
\newblock \bibinfo{journal}{\emph{Biological Cybernetics}}  \bibinfo{volume}{36} (\bibinfo{year}{1980}), \bibinfo{pages}{193--202}.
\newblock


\bibitem[Garcia and Korhonen(2018)]%
        {Garcia2018EfficientDF}
\bibfield{author}{\bibinfo{person}{Johan Garcia} {and} \bibinfo{person}{Topi Korhonen}.} \bibinfo{year}{2018}\natexlab{}.
\newblock \showarticletitle{Efficient Distribution-Derived Features for High-Speed Encrypted Flow Classification}.
\newblock \bibinfo{journal}{\emph{Proceedings of the 2018 Workshop on Network Meets AI \& ML}} (\bibinfo{year}{2018}).
\newblock
\urldef\tempurl%
\url{https://api.semanticscholar.org/CorpusID:51929029}
\showURL{%
\tempurl}


\bibitem[Haffner et~al\mbox{.}(2005)]%
        {10.1145/1080173.1080183}
\bibfield{author}{\bibinfo{person}{Patrick Haffner}, \bibinfo{person}{Subhabrata Sen}, \bibinfo{person}{Oliver Spatscheck}, {and} \bibinfo{person}{Dongmei Wang}.} \bibinfo{year}{2005}\natexlab{}.
\newblock \showarticletitle{{ACAS}: Automated Construction of Application Signatures}. In \bibinfo{booktitle}{\emph{ACM SIGCOMM MineNet'05}}. \bibinfo{pages}{197–202}.
\newblock
\showISBNx{1595930264}


\bibitem[Ho(1995)]%
        {598994}
\bibfield{author}{\bibinfo{person}{Tin~Kam Ho}.} \bibinfo{year}{1995}\natexlab{}.
\newblock \showarticletitle{Random decision forests}. In \bibinfo{booktitle}{\emph{Proceedings of 3rd International Conference on Document Analysis and Recognition}}, Vol.~\bibinfo{volume}{1}. \bibinfo{pages}{278--282 vol.1}.
\newblock
\urldef\tempurl%
\url{https://doi.org/10.1109/ICDAR.1995.598994}
\showDOI{\tempurl}


\bibitem[Holland et~al\mbox{.}(2021)]%
        {10.1145/3460120.3484758}
\bibfield{author}{\bibinfo{person}{Jordan Holland}, \bibinfo{person}{Paul Schmitt}, \bibinfo{person}{Nick Feamster}, {and} \bibinfo{person}{Prateek Mittal}.} \bibinfo{year}{2021}\natexlab{}.
\newblock \showarticletitle{New Directions in Automated Traffic Analysis}. In \bibinfo{booktitle}{\emph{ACM CCS'21}}. \bibinfo{pages}{3366–3383}.
\newblock
\showISBNx{9781450384544}


\bibitem[Horowicz et~al\mbox{.}(2022)]%
        {10.1145/3517745.3561436}
\bibfield{author}{\bibinfo{person}{Eyal Horowicz}, \bibinfo{person}{Tal Shapira}, {and} \bibinfo{person}{Yuval Shavitt}.} \bibinfo{year}{2022}\natexlab{}.
\newblock \showarticletitle{A Few Shots Traffic Classification with Mini-FlowPic Augmentations}. In \bibinfo{booktitle}{\emph{ACM IMC'22}} (Nice, France). \bibinfo{pages}{647–654}.
\newblock
\showISBNx{9781450392594}
\urldef\tempurl%
\url{https://doi.org/10.1145/3517745.3561436}
\showDOI{\tempurl}


\bibitem[Hughes(1968)]%
        {Hughes1968OnTM}
\bibfield{author}{\bibinfo{person}{Gordon~P. Hughes}.} \bibinfo{year}{1968}\natexlab{}.
\newblock \showarticletitle{On the mean accuracy of statistical pattern recognizers}.
\newblock \bibinfo{journal}{\emph{IEEE Trans. Inf. Theory}}  \bibinfo{volume}{14} (\bibinfo{year}{1968}), \bibinfo{pages}{55--63}.
\newblock
\urldef\tempurl%
\url{https://api.semanticscholar.org/CorpusID:206729491}
\showURL{%
\tempurl}


\bibitem[Jacobs et~al\mbox{.}(2022)]%
        {10.1145/3548606.3560609}
\bibfield{author}{\bibinfo{person}{Arthur~S. Jacobs}, \bibinfo{person}{Roman Beltiukov}, \bibinfo{person}{Walter Willinger}, \bibinfo{person}{Ronaldo~A. Ferreira}, \bibinfo{person}{Arpit Gupta}, {and} \bibinfo{person}{Lisandro~Z. Granville}.} \bibinfo{year}{2022}\natexlab{}.
\newblock \showarticletitle{AI/ML for Network Security: The Emperor Has No Clothes}. In \bibinfo{booktitle}{\emph{Proceedings of the 2022 ACM SIGSAC Conference on Computer and Communications Security}} (Los Angeles, CA, USA) \emph{(\bibinfo{series}{CCS '22})}. \bibinfo{publisher}{Association for Computing Machinery}, \bibinfo{address}{New York, NY, USA}, \bibinfo{pages}{1537–1551}.
\newblock
\showISBNx{9781450394505}
\urldef\tempurl%
\url{https://doi.org/10.1145/3548606.3560609}
\showDOI{\tempurl}


\bibitem[Kohavi(1995)]%
        {Kohavi1995ASO}
\bibfield{author}{\bibinfo{person}{Ron Kohavi}.} \bibinfo{year}{1995}\natexlab{}.
\newblock \showarticletitle{A Study of Cross-Validation and Bootstrap for Accuracy Estimation and Model Selection}. In \bibinfo{booktitle}{\emph{International Joint Conference on Artificial Intelligence}}.
\newblock


\bibitem[Lashkari et~al\mbox{.}(2017)]%
        {HabibiLashkari2017CharacterizationOT}
\bibfield{author}{\bibinfo{person}{Arash~Habibi Lashkari}, \bibinfo{person}{Gerard Draper-Gil}, \bibinfo{person}{Mohammad Saiful~Islam Mamun}, {and} \bibinfo{person}{Ali~A. Ghorbani}.} \bibinfo{year}{2017}\natexlab{}.
\newblock \showarticletitle{Characterization of {Tor} Traffic using Time based Features}. In \bibinfo{booktitle}{\emph{ICISSP}}.
\newblock


\bibitem[LEI(2014)]%
        {confidence}
\bibfield{author}{\bibinfo{person}{JING LEI}.} \bibinfo{year}{2014}\natexlab{}.
\newblock \showarticletitle{Classification with confidence}.
\newblock \bibinfo{journal}{\emph{Biometrika}} \bibinfo{volume}{101}, \bibinfo{number}{4} (\bibinfo{year}{2014}), \bibinfo{pages}{755--769}.
\newblock
\showISSN{00063444}
\urldef\tempurl%
\url{http://www.jstor.org/stable/43304686}
\showURL{%
\tempurl}


\bibitem[Li et~al\mbox{.}(2018)]%
        {Li2018ByteSN}
\bibfield{author}{\bibinfo{person}{Rui Li}, \bibinfo{person}{Xi Xiao}, \bibinfo{person}{Shiguang Ni}, \bibinfo{person}{Haitao Zheng}, {and} \bibinfo{person}{Shutao Xia}.} \bibinfo{year}{2018}\natexlab{}.
\newblock \showarticletitle{Byte Segment Neural Network for Network Traffic Classification}.
\newblock \bibinfo{journal}{\emph{2018 IEEE/ACM 26th International Symposium on Quality of Service (IWQoS)}} (\bibinfo{year}{2018}), \bibinfo{pages}{1--10}.
\newblock
\urldef\tempurl%
\url{https://api.semanticscholar.org/CorpusID:59232416}
\showURL{%
\tempurl}


\bibitem[Lin(1991)]%
        {jsd}
\bibfield{author}{\bibinfo{person}{Jianhua Lin}.} \bibinfo{year}{1991}\natexlab{}.
\newblock \showarticletitle{Divergence measures based on the Shannon entropy}.
\newblock \bibinfo{journal}{\emph{IEEE Trans. Inf. Theory}}  \bibinfo{volume}{37} (\bibinfo{year}{1991}), \bibinfo{pages}{145--151}.
\newblock
\urldef\tempurl%
\url{https://api.semanticscholar.org/CorpusID:12121632}
\showURL{%
\tempurl}


\bibitem[Liu et~al\mbox{.}(2019)]%
        {Liu2019FSNetAF}
\bibfield{author}{\bibinfo{person}{Chang Liu}, \bibinfo{person}{Longtao He}, \bibinfo{person}{Gang Xiong}, \bibinfo{person}{Zigang Cao}, {and} \bibinfo{person}{Zhen Li}.} \bibinfo{year}{2019}\natexlab{}.
\newblock \showarticletitle{FS-Net: A Flow Sequence Network For Encrypted Traffic Classification}.
\newblock \bibinfo{journal}{\emph{IEEE INFOCOM 2019 - IEEE Conference on Computer Communications}} (\bibinfo{year}{2019}), \bibinfo{pages}{1171--1179}.
\newblock
\urldef\tempurl%
\url{https://api.semanticscholar.org/CorpusID:86515190}
\showURL{%
\tempurl}


\bibitem[Lopez-Martin et~al\mbox{.}(2017)]%
        {8026581}
\bibfield{author}{\bibinfo{person}{Manuel Lopez-Martin}, \bibinfo{person}{Belen Carro}, \bibinfo{person}{Antonio Sanchez-Esguevillas}, {and} \bibinfo{person}{Jaime Lloret}.} \bibinfo{year}{2017}\natexlab{}.
\newblock \showarticletitle{Network Traffic Classifier With Convolutional and Recurrent Neural Networks for Internet of Things}.
\newblock \bibinfo{journal}{\emph{IEEE Access}}  \bibinfo{volume}{5} (\bibinfo{year}{2017}), \bibinfo{pages}{18042--18050}.
\newblock
\urldef\tempurl%
\url{https://doi.org/10.1109/ACCESS.2017.2747560}
\showDOI{\tempurl}


\bibitem[Lotfollahi et~al\mbox{.}(2017)]%
        {Lotfollahi2017DeepPA}
\bibfield{author}{\bibinfo{person}{Mohammad Lotfollahi}, \bibinfo{person}{Mahdi~Jafari Siavoshani}, \bibinfo{person}{Ramin Shirali~Hossein Zade}, {and} \bibinfo{person}{Mohammdsadegh Saberian}.} \bibinfo{year}{2017}\natexlab{}.
\newblock \showarticletitle{Deep packet: a novel approach for encrypted traffic classification using deep learning}.
\newblock \bibinfo{journal}{\emph{Soft Computing}}  \bibinfo{volume}{24} (\bibinfo{year}{2017}), \bibinfo{pages}{1999 -- 2012}.
\newblock
\urldef\tempurl%
\url{https://api.semanticscholar.org/CorpusID:35187639}
\showURL{%
\tempurl}


\bibitem[Moore and Papagiannaki(2005)]%
        {10.1007/978-3-540-31966-5_4}
\bibfield{author}{\bibinfo{person}{Andrew~W. Moore} {and} \bibinfo{person}{Konstantina Papagiannaki}.} \bibinfo{year}{2005}\natexlab{}.
\newblock \showarticletitle{Toward the Accurate Identification of Network Applications}. In \bibinfo{booktitle}{\emph{PAM'05}}. \bibinfo{pages}{41--54}.
\newblock
\showISBNx{978-3-540-31966-5}


\bibitem[Moore and Zuev(2005)]%
        {10.1145/1071690.1064220}
\bibfield{author}{\bibinfo{person}{Andrew~W. Moore} {and} \bibinfo{person}{Denis Zuev}.} \bibinfo{year}{2005}\natexlab{}.
\newblock \showarticletitle{Internet Traffic Classification Using Bayesian Analysis Techniques}.
\newblock \bibinfo{journal}{\emph{SIGMETRICS Perform. Eval. Rev.}} \bibinfo{volume}{33}, \bibinfo{number}{1} (\bibinfo{date}{jun} \bibinfo{year}{2005}), \bibinfo{pages}{50–60}.
\newblock
\showISSN{0163-5999}


\bibitem[Naas and Fesl(2023)]%
        {NAAS2023108945}
\bibfield{author}{\bibinfo{person}{Mohamed Naas} {and} \bibinfo{person}{Jan Fesl}.} \bibinfo{year}{2023}\natexlab{}.
\newblock \showarticletitle{A novel dataset for encrypted virtual private network traffic analysis}.
\newblock \bibinfo{journal}{\emph{Data in Brief}}  \bibinfo{volume}{47} (\bibinfo{year}{2023}), \bibinfo{pages}{108945}.
\newblock
\showISSN{2352-3409}


\bibitem[Nguyen and Armitage(2008)]%
        {4738466}
\bibfield{author}{\bibinfo{person}{Thuy~T.T. Nguyen} {and} \bibinfo{person}{Grenville Armitage}.} \bibinfo{year}{2008}\natexlab{}.
\newblock \showarticletitle{A survey of techniques for internet traffic classification using machine learning}.
\newblock \bibinfo{journal}{\emph{IEEE Communications Surveys \& Tutorials}} \bibinfo{volume}{10}, \bibinfo{number}{4} (\bibinfo{year}{2008}), \bibinfo{pages}{56--76}.
\newblock
\urldef\tempurl%
\url{https://doi.org/10.1109/SURV.2008.080406}
\showDOI{\tempurl}


\bibitem[Pearl and Reed(1920)]%
        {doi:10.1073/pnas.6.6.275}
\bibfield{author}{\bibinfo{person}{Raymond Pearl} {and} \bibinfo{person}{Lowell~J. Reed}.} \bibinfo{year}{1920}\natexlab{}.
\newblock \showarticletitle{On the Rate of Growth of the Population of the United States since 1790 and Its Mathematical Representation1}.
\newblock \bibinfo{journal}{\emph{Proceedings of the National Academy of Sciences}} \bibinfo{volume}{6}, \bibinfo{number}{6} (\bibinfo{year}{1920}), \bibinfo{pages}{275--288}.
\newblock
\urldef\tempurl%
\url{https://doi.org/10.1073/pnas.6.6.275}
\showDOI{\tempurl}


\bibitem[Rezaei and Liu(2020)]%
        {rezaei2020achieve}
\bibfield{author}{\bibinfo{person}{Shahbaz Rezaei} {and} \bibinfo{person}{Xin Liu}.} \bibinfo{year}{2020}\natexlab{}.
\newblock \bibinfo{title}{How to Achieve High Classification Accuracy with Just a Few Labels: A Semi-supervised Approach Using Sampled Packets}.
\newblock
\newblock
\showeprint[arxiv]{1812.09761}~[cs.NI]
\urldef\tempurl%
\url{https://arxiv.org/abs/1812.09761}
\showURL{%
\tempurl}


\bibitem[Roughan et~al\mbox{.}(2004)]%
        {Roughan2004ClassofserviceMF}
\bibfield{author}{\bibinfo{person}{Matthew Roughan}, \bibinfo{person}{Subhabrata Sen}, \bibinfo{person}{Oliver Spatscheck}, {and} \bibinfo{person}{Nick~G. Duffield}.} \bibinfo{year}{2004}\natexlab{}.
\newblock \showarticletitle{Class-of-service mapping for QoS: a statistical signature-based approach to IP traffic classification}. In \bibinfo{booktitle}{\emph{ACM/SIGCOMM Internet Measurement Conference}}.
\newblock


\bibitem[Salman et~al\mbox{.}(2020)]%
        {survey2}
\bibfield{author}{\bibinfo{person}{Ola Salman}, \bibinfo{person}{Imad~H Elhajj}, \bibinfo{person}{Ayman Kayssi}, {and} \bibinfo{person}{Ali Chehab}.} \bibinfo{year}{2020}\natexlab{}.
\newblock \showarticletitle{A review on machine learning--based approaches for Internet traffic classification}.
\newblock \bibinfo{journal}{\emph{Annals of Telecommunications}} \bibinfo{volume}{75}, \bibinfo{number}{11} (\bibinfo{year}{2020}), \bibinfo{pages}{673--710}.
\newblock


\bibitem[Shafiq et~al\mbox{.}(2018)]%
        {Shafiq2018AML}
\bibfield{author}{\bibinfo{person}{Muhammad Shafiq}, \bibinfo{person}{Xiangzhan Yu}, \bibinfo{person}{Ali~Kashif Bashir}, \bibinfo{person}{Hassan~Nazeer Chaudhry}, {and} \bibinfo{person}{Dawei Wang}.} \bibinfo{year}{2018}\natexlab{}.
\newblock \showarticletitle{A machine learning approach for feature selection traffic classification using security analysis}.
\newblock \bibinfo{journal}{\emph{The Journal of Supercomputing}}  \bibinfo{volume}{74} (\bibinfo{year}{2018}), \bibinfo{pages}{4867--4892}.
\newblock


\bibitem[Shapira and Shavitt(2019)]%
        {shapira2019flowpic}
\bibfield{author}{\bibinfo{person}{Tal Shapira} {and} \bibinfo{person}{Yuval Shavitt}.} \bibinfo{year}{2019}\natexlab{}.
\newblock \showarticletitle{Flowpic: Encrypted internet traffic classification is as easy as image recognition}. In \bibinfo{booktitle}{\emph{IEEE Infocom Workshops}}. \bibinfo{pages}{680--687}.
\newblock


\bibitem[Sharafaldin et~al\mbox{.}(2018)]%
        {Sharafaldin2018TowardGA}
\bibfield{author}{\bibinfo{person}{Iman Sharafaldin}, \bibinfo{person}{Arash~Habibi Lashkari}, {and} \bibinfo{person}{Ali~A. Ghorbani}.} \bibinfo{year}{2018}\natexlab{}.
\newblock \showarticletitle{Toward Generating a New Intrusion Detection Dataset and Intrusion Traffic Characterization}. In \bibinfo{booktitle}{\emph{ICISSP}}.
\newblock


\bibitem[Sivanathan et~al\mbox{.}(2019)]%
        {Sivanathan2019ClassifyingID}
\bibfield{author}{\bibinfo{person}{Arunan Sivanathan}, \bibinfo{person}{Hassan~Habibi Gharakheili}, \bibinfo{person}{Franco Loi}, \bibinfo{person}{Adam Radford}, \bibinfo{person}{Chamith Wijenayake}, \bibinfo{person}{Arun Vishwanath}, {and} \bibinfo{person}{Vijay Sivaraman}.} \bibinfo{year}{2019}\natexlab{}.
\newblock \showarticletitle{Classifying IoT Devices in Smart Environments Using Network Traffic Characteristics}.
\newblock \bibinfo{journal}{\emph{IEEE Transactions on Mobile Computing}}  \bibinfo{volume}{18} (\bibinfo{year}{2019}), \bibinfo{pages}{1745--1759}.
\newblock
\urldef\tempurl%
\url{https://api.semanticscholar.org/CorpusID:70082542}
\showURL{%
\tempurl}


\bibitem[van Ede et~al\mbox{.}(2020)]%
        {Ede2020FlowPrintSM}
\bibfield{author}{\bibinfo{person}{Thijs van Ede}, \bibinfo{person}{Riccardo Bortolameotti}, \bibinfo{person}{Andrea Continella}, \bibinfo{person}{Jingjing Ren}, \bibinfo{person}{Daniel~J. Dubois}, \bibinfo{person}{Martina Lindorfer}, \bibinfo{person}{David~R. Choffnes}, \bibinfo{person}{Maarten van Steen}, {and} \bibinfo{person}{Andreas Peter}.} \bibinfo{year}{2020}\natexlab{}.
\newblock \showarticletitle{FlowPrint: Semi-Supervised Mobile-App Fingerprinting on Encrypted Network Traffic}.
\newblock \bibinfo{journal}{\emph{Proceedings 2020 Network and Distributed System Security Symposium}} (\bibinfo{year}{2020}).
\newblock
\urldef\tempurl%
\url{https://api.semanticscholar.org/CorpusID:211265114}
\showURL{%
\tempurl}


\bibitem[Wang et~al\mbox{.}(2018)]%
        {8171733}
\bibfield{author}{\bibinfo{person}{Wei Wang}, \bibinfo{person}{Yiqiang Sheng}, \bibinfo{person}{Jinlin Wang}, \bibinfo{person}{Xuewen Zeng}, \bibinfo{person}{Xiaozhou Ye}, \bibinfo{person}{Yongzhong Huang}, {and} \bibinfo{person}{Ming Zhu}.} \bibinfo{year}{2018}\natexlab{}.
\newblock \showarticletitle{HAST-IDS: Learning Hierarchical Spatial-Temporal Features Using Deep Neural Networks to Improve Intrusion Detection}.
\newblock \bibinfo{journal}{\emph{IEEE Access}}  \bibinfo{volume}{6} (\bibinfo{year}{2018}), \bibinfo{pages}{1792--1806}.
\newblock
\urldef\tempurl%
\url{https://doi.org/10.1109/ACCESS.2017.2780250}
\showDOI{\tempurl}


\bibitem[Wang et~al\mbox{.}(2017a)]%
        {8004872}
\bibfield{author}{\bibinfo{person}{Wei Wang}, \bibinfo{person}{Ming Zhu}, \bibinfo{person}{Jinlin Wang}, \bibinfo{person}{Xuewen Zeng}, {and} \bibinfo{person}{Zhongzhen Yang}.} \bibinfo{year}{2017}\natexlab{a}.
\newblock \showarticletitle{End-to-end encrypted traffic classification with one-dimensional convolution neural networks}. In \bibinfo{booktitle}{\emph{2017 IEEE International Conference on Intelligence and Security Informatics (ISI)}}. \bibinfo{pages}{43--48}.
\newblock
\urldef\tempurl%
\url{https://doi.org/10.1109/ISI.2017.8004872}
\showDOI{\tempurl}


\bibitem[Wang et~al\mbox{.}(2017b)]%
        {Wang2017EndtoendET}
\bibfield{author}{\bibinfo{person}{Wei Wang}, \bibinfo{person}{Ming Zhu}, \bibinfo{person}{Jinlin Wang}, \bibinfo{person}{Xuewen Zeng}, {and} \bibinfo{person}{Zhongzhen Yang}.} \bibinfo{year}{2017}\natexlab{b}.
\newblock \showarticletitle{End-to-end encrypted traffic classification with one-dimensional convolution neural networks}.
\newblock \bibinfo{journal}{\emph{2017 IEEE International Conference on Intelligence and Security Informatics (ISI)}} (\bibinfo{year}{2017}), \bibinfo{pages}{43--48}.
\newblock
\urldef\tempurl%
\url{https://api.semanticscholar.org/CorpusID:3720713}
\showURL{%
\tempurl}


\bibitem[Wang et~al\mbox{.}(2023)]%
        {TaTic}
\bibfield{author}{\bibinfo{person}{Yipeng Wang}, \bibinfo{person}{Huijie He}, \bibinfo{person}{Yingxu Lai}, {and} \bibinfo{person}{Alex~X. Liu}.} \bibinfo{year}{2023}\natexlab{}.
\newblock \showarticletitle{A Two-Phase Approach to Fast and Accurate Classification of Encrypted Traffic}.
\newblock \bibinfo{journal}{\emph{IEEE/ACM Transactions on Networking}} \bibinfo{volume}{31}, \bibinfo{number}{3} (\bibinfo{year}{2023}), \bibinfo{pages}{1071--1086}.
\newblock
\urldef\tempurl%
\url{https://doi.org/10.1109/TNET.2022.3209979}
\showDOI{\tempurl}


\bibitem[Williams et~al\mbox{.}(2006)]%
        {article}
\bibfield{author}{\bibinfo{person}{Nigel Williams}, \bibinfo{person}{Sebastian Zander}, {and} \bibinfo{person}{Grenville Armitage}.} \bibinfo{year}{2006}\natexlab{}.
\newblock \showarticletitle{A preliminary performance comparison of five machine learning algorithms for practical {IP} traffic flow classification}.
\newblock \bibinfo{journal}{\emph{SIGCOMM Comput. Commun. Rev.}}  \bibinfo{volume}{36} (\bibinfo{date}{10} \bibinfo{year}{2006}), \bibinfo{pages}{5--16}.
\newblock
\urldef\tempurl%
\url{https://doi.org/10.1145/1163593.1163596}
\showDOI{\tempurl}


\bibitem[Xu et~al\mbox{.}(2023)]%
        {FastTraffic}
\bibfield{author}{\bibinfo{person}{Yuwei Xu}, \bibinfo{person}{Jie Cao}, \bibinfo{person}{Kehui Song}, \bibinfo{person}{Qiao Xiang}, {and} \bibinfo{person}{Guang Cheng}.} \bibinfo{year}{2023}\natexlab{}.
\newblock \showarticletitle{FastTraffic: A lightweight method for encrypted traffic fast classification}.
\newblock \bibinfo{journal}{\emph{Computer Networks}}  \bibinfo{volume}{235} (\bibinfo{year}{2023}), \bibinfo{pages}{109965}.
\newblock
\showISSN{1389-1286}
\urldef\tempurl%
\url{https://doi.org/10.1016/j.comnet.2023.109965}
\showDOI{\tempurl}


\bibitem[Yang et~al\mbox{.}(2021)]%
        {9585567}
\bibfield{author}{\bibinfo{person}{Lixuan Yang}, \bibinfo{person}{Alessandro Finamore}, \bibinfo{person}{Feng Jun}, {and} \bibinfo{person}{Dario Rossi}.} \bibinfo{year}{2021}\natexlab{}.
\newblock \showarticletitle{Deep Learning and Zero-Day Traffic Classification: Lessons Learned From a Commercial-Grade Dataset}.
\newblock \bibinfo{journal}{\emph{IEEE TNSM}} \bibinfo{volume}{18}, \bibinfo{number}{4} (\bibinfo{year}{2021}), \bibinfo{pages}{4103--4118}.
\newblock
\urldef\tempurl%
\url{https://doi.org/10.1109/TNSM.2021.3122940}
\showDOI{\tempurl}


\bibitem[Zhao et~al\mbox{.}(2021)]%
        {survey1}
\bibfield{author}{\bibinfo{person}{Jingjing Zhao}, \bibinfo{person}{Xuyang Jing}, {and} \bibinfo{person}{Witold Pedrycz}.} \bibinfo{year}{2021}\natexlab{}.
\newblock \showarticletitle{Network traffic classification for data fusion: A survey}.
\newblock \bibinfo{journal}{\emph{Information Fusion}}  \bibinfo{volume}{72} (\bibinfo{date}{02} \bibinfo{year}{2021}).
\newblock
\urldef\tempurl%
\url{https://doi.org/10.1016/j.inffus.2021.02.009}
\showDOI{\tempurl}


\bibitem[Zou et~al\mbox{.}(2018)]%
        {Zou2018EncryptedTC}
\bibfield{author}{\bibinfo{person}{Zhuang Zou}, \bibinfo{person}{Jingguo Ge}, \bibinfo{person}{Hongbo Zheng}, \bibinfo{person}{Yulei Wu}, \bibinfo{person}{Chunjing Han}, {and} \bibinfo{person}{Zhongjiang Yao}.} \bibinfo{year}{2018}\natexlab{}.
\newblock \showarticletitle{Encrypted Traffic Classification with a Convolutional Long Short-Term Memory Neural Network}.
\newblock \bibinfo{journal}{\emph{2018 IEEE 20th International Conference on High Performance Computing and Communications; IEEE 16th International Conference on Smart City; IEEE 4th International Conference on Data Science and Systems (HPCC/SmartCity/DSS)}} (\bibinfo{year}{2018}), \bibinfo{pages}{329--334}.
\newblock
\urldef\tempurl%
\url{https://api.semanticscholar.org/CorpusID:59233070}
\showURL{%
\tempurl}


\end{thebibliography}
